 \documentclass[authoryear,preprint,12pt]{elsarticle}
\usepackage{lineno}
\usepackage[table]{xcolor}

\usepackage{amssymb}
\usepackage{array}
\newcolumntype{C}[1]{>{\centering\let\newline\\\arraybackslash\hspace{0pt}}m{#1}}
\usepackage{longtable}
\usepackage{pdflscape}

 \usepackage{amsthm}
 \usepackage{amsmath}

\usepackage{booktabs}
\usepackage{tabularx}

\usepackage{multirow}
\usepackage{caption}
\usepackage{subcaption}
\usepackage{tabularx}
\usepackage{bm}

\usepackage{pgf, tikz} % per grafici direttamente con latex
\usetikzlibrary{arrows,automata,fit}
\usetikzlibrary{shapes}

\usepackage{booktabs}
\usepackage{hyperref}
\journal{}

\begin{document}
%\linenumbers
\begin{frontmatter}



\title{Disentangling Spatial and Structural Drivers of Housing Prices through Bayesian Networks: A Case Study of Madrid, Barcelona, and Valencia}

\author[label1]{Alvaro Garcia Murga}
 \author[label1]{Manuele Leonelli}
 \affiliation[label1]{organization={School of Science and Technology, IE University, Madrid}, country = {Spain}}

%%
%% \affiliation[label2,label3]{organization={},

\begin{abstract}
Understanding how housing prices respond to spatial accessibility, structural attributes, and typological distinctions is central to contemporary urban research and policy. In cities marked by affordability stress and market segmentation, models that offer both predictive capability and interpretive clarity are increasingly needed. This study applies discrete Bayesian networks to model residential price formation across Madrid, Barcelona, and Valencia using over 180,000 geo-referenced housing listings. The resulting probabilistic structures reveal distinct city-specific logics. Madrid exhibits amenity-driven stratification, Barcelona emphasizes typology and classification, while Valencia is shaped by spatial and structural fundamentals. By enabling joint inference, scenario simulation, and sensitivity analysis within a transparent framework, the approach advances housing analytics toward models that are not only accurate but actionable, interpretable, and aligned with the demands of equitable urban governance.

\end{abstract}



%\item Discrete Bayesian networks for modeling housing prices in Spanish cities
%\item Integration of structural, spatial, and typological features from listing data
%\item Identification of city-specific price drivers through probabilistic inference
%\item Scenario-based simulation of market segments across urban profiles
%\item Interpretability and policy relevance for housing equity and planning


\end{frontmatter}


\section{Introduction}
Understanding the determinants of housing prices is central to urban policy, planning, and inequality research. In cities facing affordability stress, spatial segregation, and speculative investment pressure, housing prices both reflect and reproduce structural dynamics in the built environment \citep{leal2025decoding, martinez2020house}. As housing systems grow more fragmented and spatially uneven, the demand for robust, data-driven models to inform regulation, taxation, and planning has intensified. Spain exemplifies these pressures. Its major urban centers (Madrid, Barcelona, and Valencia) exhibit persistent demand–supply mismatches, pronounced spatial inequality, and layered submarkets shaped by infrastructure, typology, and legacy regulation \citep{ kenyon2024intra, rey2023mlseg}.

Despite growing data availability, most modeling tools used in practice remain either overly rigid or opaque. Hedonic regressions offer interpretability but often struggle with non-linearities, interactions, and spatial heterogeneity. Conversely, machine learning models offer predictive gains but are difficult to audit and explain, particularly in high-stakes contexts such as property taxation or housing equity assessments \citep{lorenz2023interpretable, rudin2019stop}. While post hoc explainability tools (e.g. SHAP or LIME) have improved transparency, they remain approximate by design and offer limited insight into underlying causal or structural logic \citep{gunes2024model, trindade2024impacts}.

In this study, we apply discrete Bayesian networks (BNs) as a fully interpretable modeling framework to estimate housing prices across three major Spanish cities. Leveraging a large, geo-referenced dataset from the Idealista platform comprising over 180,000 residential listings, we develop intuitive models that encode structural, spatial, and typological dependencies in a transparent and policy-relevant form. BNs allow for joint inference, scenario-based simulation, and robust sensitivity analysis, supporting both descriptive insight and decision-making. 

By capturing city-specific pricing structures in an interpretable, probabilistic framework, this study offers a new lens on how housing markets operate across space, structure, and policy context. It demonstrates how BNs can be deployed not only for accurate modeling, but for generating actionable insight in one of the most socially consequential domains of applied urban analytics.

\section{Literature Review}

\subsection{The Spanish Housing Market}

The Spanish housing market has experienced deep structural shifts over the past two decades. Between 1997 and 2007, Spain underwent one of the most intense housing booms in the OECD, driven by deregulated mortgage lending and pro-ownership policies. The national homeownership rate rose from 45\% in 1950 to 85\% by 2007, among the highest in Europe \citep{lopez2011spanish}. The 2008 financial crisis led to a sharp correction in prices and construction activity, with more than 378,000 evictions recorded between 2008 and 2014, the highest number in Europe \citep{kenyon2024intra}.

Since 2014, the sector has re-entered an expansionary phase, yet structural constraints persist. Housing demand is estimated at over 250,000 new units per year, while completions remain consistently below 100,000 \citep{lucasfox2024}. This mismatch is expected to continue for at least 15 years. New housing sales have increased by 24\% compared to the previous decade, yet approvals for new projects have dropped by 53\% over the same period \citep{tinsa2024}. Buyer sentiment remains strong: 70\% of Spaniards expect prices to rise further, and 59\% cite proximity to services as a key factor when choosing a home \citep{ingspain2023}.

City-level dynamics vary. In Madrid, price stratification reflects spatial segmentation around corridors like Castellana and Gran Vía. Barcelona emphasizes typology and dwelling classification, while Valencia displays a more structurally driven logic based on building age, size, and accessibility \citep{rey2023mlseg, rey2024improving}. These patterns are reinforced by rising spatial inequality: between 2010 and 2015, average prices in Madrid declined, yet central districts such as Salamanca and Chamartín retained over half the city’s housing wealth, while areas like Usera and Villaverde remained under-valued \citep{kenyon2024intra}. Public housing policies have further exacerbated this divide, with protected housing now accounting for just 2.8\% of the national stock \citep{martinez2020house}.

Idealista has emerged as the de facto data source for market analysis, covering 80–91\% of listings in Spain’s largest cities \citep{rey2024improving}. However, its spatial completeness and representativeness remain open to debate \citep{larraz2011expert}. Comparative research across Europe (such as in Lisbon, where rental regimes differ across segments) also supports a move toward more spatially disaggregated and typology-aware housing policy \citep{leal2025decoding}.

\subsection{Structural Determinants of Housing Prices}

Structural features remain central to housing price formation. Surface area, floor level, room count, age, and presence of amenities such as elevators, parking, or terraces are among the most cited determinants \citep{alfaro2020fully, baldominos2018identifying, rico2021machine, taltavull2003determinants}. In high-end markets like Salamanca (Madrid), structural characteristics explain much of the variance in asking price, though location exerts independent influence even after controlling for these factors \citep{baldominos2018identifying}.

Their effects, however, are not uniform. In Alicante, income and location dominate in upper price quantiles, while housing age and structural quality have stronger effects at the lower end of the market \citep{rico2021machine}. Smaller municipalities show higher pricing volatility and weaker structure-price correlations, a result of thin markets and asymmetric information \citep{alfaro2020fully}.

International studies confirm the importance of structure but note the amplifying or attenuating role of local expectations and socio-cultural preferences. In Tehran, for instance, access to sports fields and water infrastructure significantly increased prices, while gas station proximity depressed values, particularly in affluent districts \citep{rajaei2024tehran}. In Rome, elevator access, structural condition, and floor level increased prices, while garden presence had a negative impact, possibly due to maintenance burden or location bias \citep{guarini2025intrinsic}.

Typology is especially salient in Southern European markets. Listings for “piso señorial”, penthouses, or duplexes in Barcelona command premiums not purely for structural advantages but for their symbolic capital and lifestyle positioning \citep{rey2023mlseg}. This effect is mirrored in Lisbon, where new contracts reflect market dynamics while older ones are shaped by legacy regulation \citep{leal2025decoding}.

Explainable AI models have confirmed the importance of structural attributes in high-dimensional settings. In Seoul, floor level and building year ranked among the strongest predictors of rental value, though their marginal effect varied spatially and across tenure types \citep{kim2024seoul}. These findings support joint modeling of structure with context-specific factors such as tenure, local regulation, and neighborhood characteristics.

\subsection{Spatial Determinants and Market Segmentation}

Location continues to be the strongest predictor of price variation across Spanish cities \citep{baldominos2018identifying, rey2023mlseg, rey2024geo}. Traditional proxies such as Euclidean distance or administrative district often fail to capture the functional logic of urban space \citep{alfaro2020fully, rico2021machine}. Consequently, recent studies advocate for polycentric and accessibility-aware models that reflect overlapping influences from transport, amenities, and neighborhood identity \citep{heyman2018location}.

Spatial autocorrelation is a persistent feature. Intra-urban price clustering reaches 47.9\% in Madrid, 40.3\% in Valencia, and 32.2\% in Barcelona \citep{rey2023mlseg}. These patterns align with corridor effects (e.g. Castellana in Madrid) and are consistent with international evidence from cities like Lisbon and Seoul \citep{leal2025decoding, kim2024seoul}. Submarkets, defined as bounded zones with internally consistent price logic, have been documented across Spanish metros \citep{chasco2018scan, royuela2013housi}, with segmentation shaped by infrastructure, social class, and urban form.

Amenities such as green space and transit access exhibit robust price premiums. In Castellón, housing prices fall with distance to parks \citep{morancho2003hedonic}; in Seville, green space access drives cross-neighborhood differentials \citep{ramirez2022influence}; in Shenzhen, green visibility adds value across diverse income segments \citep{wu2015impact}. A recent review confirms these findings across multiple contexts, noting that access to green amenities and metro proximity consistently ranks among the top predictors in hybrid price models \citep{moreno2025review}. Studies in Shanghai and Seoul also identify nonlinear thresholds in accessibility effects, e.g. subway distance beyond 20 minutes sharply reduces housing value \citep{dou2023neighborhoods, kim2024seoul}.

Segmentation is not merely spatial but institutional. In Lisbon, old rental contracts and social housing generate spatial discontinuities in value, even within walkable central zones \citep{leal2025decoding}. In Madrid, similar disparities emerge between public housing clusters and gentrified districts. The spatial logic of price is further complicated by financialisation: foreign capital and tourism investment have concentrated in central zones, amplifying spatial inequality and displacing long-term residents \citep{martinez2020house}.

\subsection{Modeling Approaches in Housing Price Estimation}

The estimation of housing prices has long relied on traditional statistical models, particularly hedonic pricing models grounded in the work of \citet{rosen1974hedonic}. These models offer interpretability and a clear economic rationale, enabling transparent decomposition of price into structural and locational attributes. However, they are limited in their ability to capture complex interactions, spatial dependencies, and non-linearities \citep{bourassa2025reflections}. Multicollinearity and functional form misspecification remain persistent challenges \citep{alfaro2020fully,moreno2025review}.

In response, recent years have seen a marked shift toward machine learning  approaches, particularly tree-based algorithms such as Random Forests and XGBoost, which offer superior predictive performance across a range of real estate markets \citep{gunes2024model,moreno2025review}. These methods are better suited to high-dimensional and heterogeneous datasets, particularly those derived from listing platforms or enriched with contextual open data \citep{trindade2024impacts}. Yet, their adoption has raised critical concerns over transparency, especially in high-stakes contexts such as property taxation, public planning, or regulatory pricing \citep{bourassa2025reflections,lorenz2023interpretable}.

To bridge the gap between predictive power and interpretability, the field has increasingly turned to explainable AI. Tools such as SHAP, LIME, Accumulated Local Effects, and Partial Dependence Plots allow post hoc interpretation of black-box models, recovering insights into variable contributions at both global and local levels \citep{gunes2024model, lorenz2023interpretable, trindade2024impacts}. These methods have clarified how structural and spatial features interact nonlinearly, and how amenity effects vary by context, e.g. the marginal impact of balconies depending on building age or the saturation of price gains from transit access \citep{dou2023neighborhoods, kim2024seoul}.

However, post hoc interpretability remains fundamentally approximate. As argued by \citet{rudin2019stop}, explanations generated for opaque models can be misleading and unfaithful to the original model logic, particularly in high-stakes contexts where interpretability is not optional but essential. \citet{rudin2019stop} calls for a shift away from black-box models altogether, advocating instead for inherently interpretable methods whose structure transparently reflects the data-generating process.

\subsection{Gaps and Contributions}

This paper responds to the growing demand for interpretable models in housing price analysis. While recent work has increasingly focused on post hoc explainability for black-box predictors, inherently transparent modeling approaches remain rare in urban real estate research. Bayesian networks (BNs), though widely adopted in fields such as environmental modeling, risk assessment, and energy systems \citep{borunda2016bayesian, kaikkonen2020bayesian}, have seen little application in real estate contexts, particularly in multi-city, spatially structured settings.

Only a handful of BN-based studies address housing price modeling \citep{constantinou2017future, liu2018bayesian, sevinc2021}, and most are limited in scope or geography. Broader adoption has occurred in adjacent urban domains such as underground planning \citep{xu2023} and mobility systems \citep{fusco2003}, and BNs have demonstrated consistent advantages in interpretability and stakeholder trust in complex environmental models \citep{kelly2013}.

We contribute to filling this gap by presenting one of the first multi-city, data-driven BN models for housing price formation. Using a rich, geo-referenced dataset of housing listings, we construct transparent graphical models for Madrid, Barcelona, and Valencia. These models enable probabilistic inference over the joint effects of structural and spatial variables, support local and global sensitivity analysis, and offer policy-relevant scenario simulation. In doing so, we demonstrate the suitability of BNs for real estate analytics, providing a framework that balances clarity and complexity in one of the most spatially and socially relevant domains of applied modeling.

\section{Materials and Methods}

\subsection{Data}

This study uses a geo-referenced dataset of residential property listings from Spain’s three largest cities: Madrid, Barcelona, and Valencia. The data originate from \textit{Idealista}, Spain’s leading real estate platform, and are freely distributed in the \texttt{idealista18} R package \citep{rey2024geo}. Idealista is the dominant online portal for real estate listings in Spain, hosting both private and professional advertisements. The dataset reflects this wide coverage, including over 189{,}000 listings from the year 2018 (94{,}815 in Madrid, 61{,}486 in Barcelona, and 33{,}622 in Valencia). These totals represent approximately 6–8\% of the housing stock in each city.

The dataset includes a rich set of variables describing each property's physical characteristics, cadastral information, and geographical coordinates. For each property, the data contain latitude and longitude, which were slightly perturbed  to preserve privacy, and key structural attributes, such as constructed area, number of rooms and bathrooms, and indicators for the presence of amenities like terrace, lift, air conditioning, swimming pool, and parking space. From the Spanish cadastre, the year of construction (used to compute building age), the maximum number of building floors, the number of dwellings, and a cadastral quality index are retained \citep{rey2024geo}. To support spatial analysis, the original dataset includes distances to key reference points such as the city center, the nearest metro station, and a major avenue in each city. We further enriched the spatial layer to better capture accessibility patterns and urban context (see Section~\ref{sec:preprocessing}).

The asking price per square meter, computed from total price and constructed area, serves as the target variable in our analysis. In the original dataset, prices were slightly masked by applying a small random perturbation and rounding to the nearest thousand euros, as described in \citet{rey2024geo}.

Open datasets of this kind are rare. One of the only other examples is a dataset of property transactions from South Korea \citep{song2021hedonic}. Despite its richness, \texttt{idealista18} has seen limited use \citep{rey2024improving,lopez2021intraurban,rey2023mlseg}, often through subsets or extended versions. While the data refer to the year 2018 and may not reflect current market conditions, it remains well suited to structural analysis. Since our focus is on uncovering the relationships between property characteristics and price formation, the dataset offers a valuable and timely resource.

\begin{figure}
\centering
\includegraphics[width=0.33\textwidth]{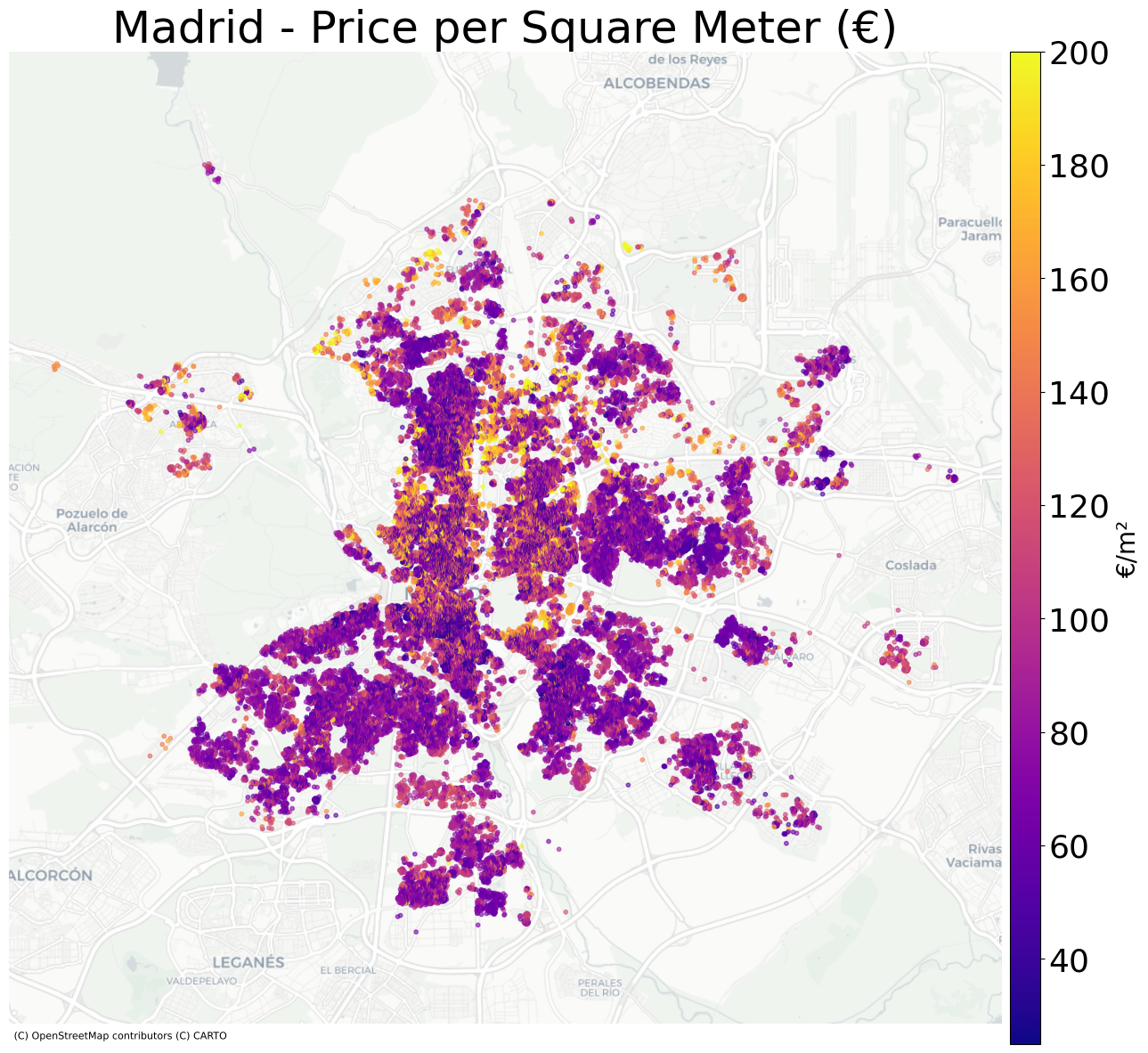}\includegraphics[width=0.33\textwidth]{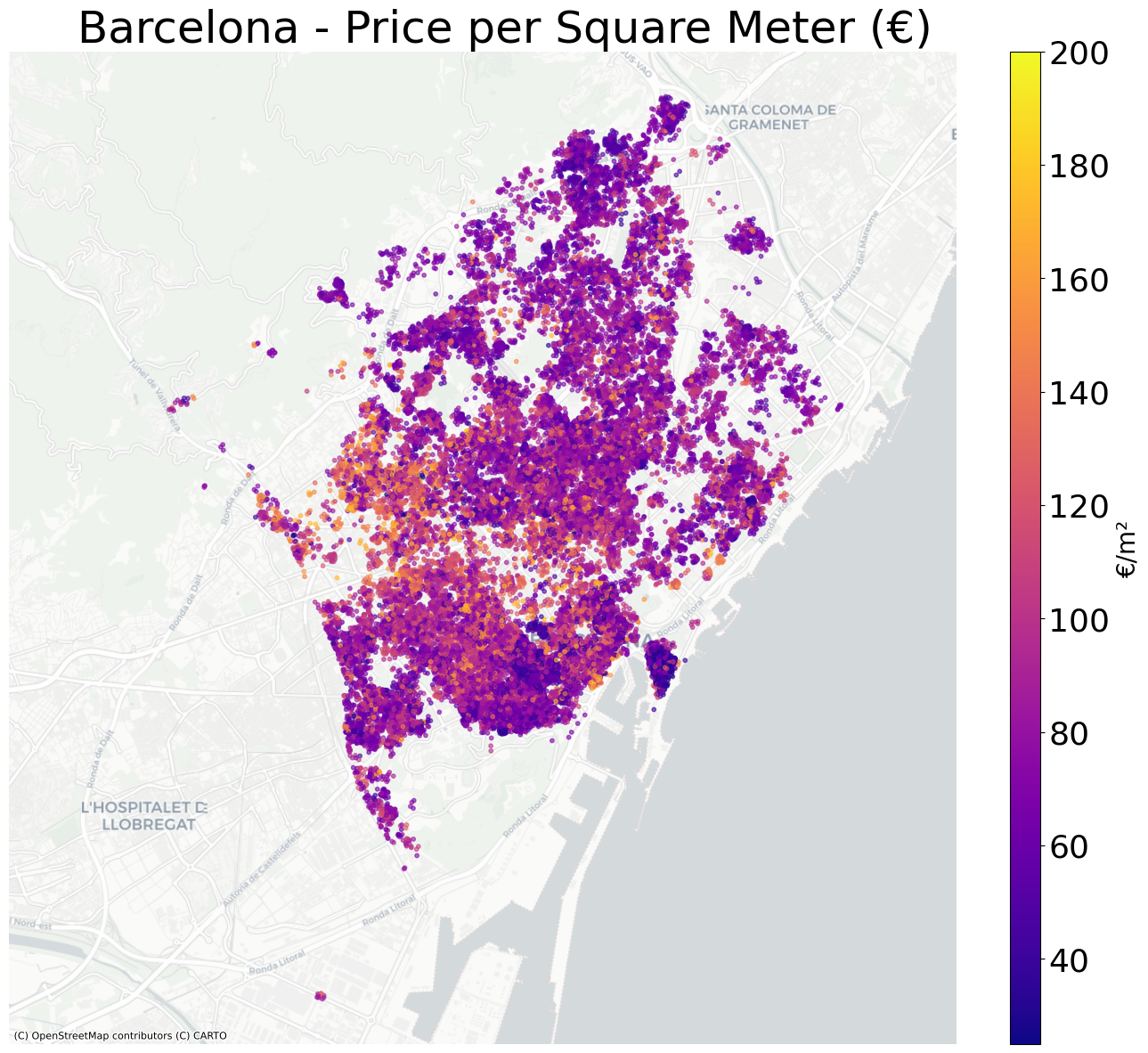}
\includegraphics[width=0.33\textwidth]{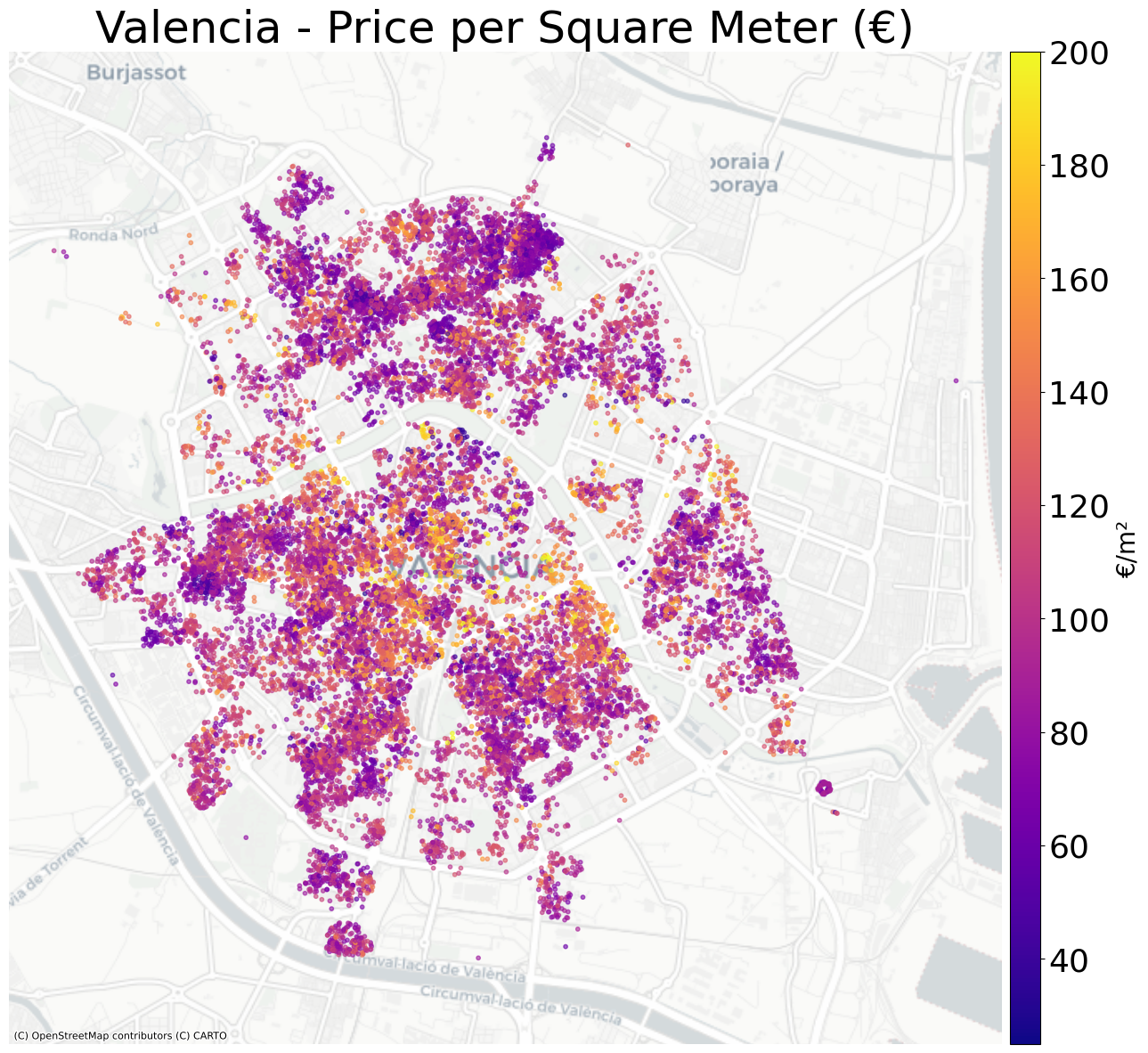}
\caption{Spatial distribution of asking prices per square meter for residential listings in Madrid, Barcelona, and Valencia.}
\label{fig:map_pricesqm}
\end{figure}

A full list of the variables used in the analysis, including definitions and value categories, is provided in Table~\ref{tab:variables}. Figure~\ref{fig:map_pricesqm} displays the spatial distribution of asking prices per square meter across Madrid, Barcelona, and Valencia. These plots reveal clear intra-city heterogeneity, with higher values concentrated in central and selected peripheral neighborhoods. The observed spatial structure motivates the inclusion of detailed accessibility measures in the modeling framework (see Section~\ref{sec:preprocessing}).

\subsection{Data Preprocessing and Feature Engineering}
\label{sec:preprocessing}

To prepare the data for BN modeling, we implemented a comprehensive data cleaning and feature engineering pipeline, applied uniformly to the listings from Madrid, Barcelona, and Valencia. This ensured consistent input across cities and aligned with the requirements of discrete probabilistic modeling.

\paragraph{Initial data cleaning}
The dataset was cleaned by converting data types, standardizing column names, and removing non-informative identifiers.
 Derived variables that could induce leakage, such as total price, were dropped. We also eliminated listings with implausible or missing values, including zero rooms or bathrooms, floor levels exceeding building height, invalid cadastral quality scores, and incorrect construction years.

\paragraph{Deduplication and spatial validation}
Duplicate listings, arising when multiple agents advertise the same property, were removed based on a combination of structural and amenity characteristics. Properties were retained only if their coordinates fell within the official administrative boundaries, defined using public \texttt{.geojson} files and matched spatially using \texttt{GeoPandas}.

\paragraph{Outlier removal}
To improve robustness, an interquartile range (IQR)-based filter was applied to key numeric variables, excluding observations beyond 2.0 IQR from the 25th and 75th percentiles. This removed extreme outliers while preserving relevant heterogeneity in property attributes.

\paragraph{Structural feature engineering}
To reduce sparsity and enhance interpretability, several new structural variables were created. Three binary indicators (\texttt{STUDIO}, \texttt{DUPLEX}, and \texttt{PENTHOUSE}) were collapsed into a single categorical variable \texttt{TYPE}. Likewise, the three condition indicators for property status (e.g., new construction, second-hand with or without renovation) were unified into a variable \texttt{CONDITION}. Building age was computed from the cadastral construction year.

\begin{table}
\footnotesize
\centering
\scalebox{0.65}{
\renewcommand{\arraystretch}{1.2}
\begin{tabular}{p{2.5cm}p{2.5cm}p{6.8cm}p{7.3cm}}
\toprule
\textbf{Group} & \textbf{Variable} & \textbf{Description} & \textbf{Levels} \\
\midrule
\multirow{10}{*}{\textbf{Structural}} & AGE & Age of the building & New Development, Modern, Mid-Age, Historic \\
 & AREA & Constructed area in m² & Small, Medium, Large, Luxury \\
 & HEIGHT & Maximum building height & Low-Rise, Mid-Rise, High-Rise, Skyscraper \\
 & FLOOR & Floor level in the building & Lower, Mid, Upper, Top \\
 & ROOMS & Number of rooms (binned) & Few, Moderate, Many \\
 & BATHS & Number of bathrooms (binned) & Few, Moderate, Many \\
 & DENSITY & Number of dwellings in the building & Low, Medium, High, Very High-Density \\
 & TYPE & Property type & Studio, Duplex, Penthouse, Standard \\
 & QUALITY & Cadastral building quality index & Low Value, Moderate Value, High Value, Very High Value \\
 & CONDITION & Property condition & New Construction, Second Hand Renovation, Second Hand Good Condition \\
 \midrule
\multirow{7}{*}{\textbf{Spatial}} & CENTRE & Distance to city centre & Very Near, Near, Medium, Far \\
 & GREEN & Distance to green space & Very Near, Near, Medium, Far \\
 & METRO & Distance to metro station & Very Near, Near, Medium, Far \\
 & MARKET & Distance to supermarket & Very Near, Near, Medium, Far \\
 & STREET1 & Distance to primary avenue (e.g., Gran Vía, Diagonal, Reino Valencia) & Very Near, Near, Medium, Far \\
 & STREET2 & Distance to secondary avenue (e.g., Castellana, Ramblas, Blasco Ibáñez) & Very Near, Near, Medium, Far \\
 & NBHD & Neighborhood frequency (quartile rank) & Most Common, Frequent, Less Frequent, Rare \\
 \midrule
\multirow{9}{*}{\textbf{Amenities}} & AC & Air conditioning available & Yes, No \\
 & LIFT & Lift present & Yes, No \\
 & PRKG & Parking space available & Yes, No \\
 & POOL & Swimming pool present & Yes, No \\
 & GARDEN & Garden present & Yes, No \\
 & TRRC & Terrace present & Yes, No \\
 & STRG & Storage room available & Yes, No \\
 & WRDRB & Built-in wardrobes & Yes, No \\
 & DMAN & Doorman service present & Yes, No \\
 \midrule
\multirow{1}{*}{\textbf{Target}} & PRICE & Asking price per m² (binned) & Very Low, Low, Medium Low, Medium High, High, Luxury \\
\bottomrule
\end{tabular}}
\caption{Grouped variables used in the Bayesian network model for Valencia, with PRICE representing binned price per square meter.}
\label{tab:variables}
\end{table}

\paragraph{Spatial feature engineering}
Our analysis substantially enriches the spatial representation of each listing by incorporating manually geocoded urban features (using sources such as Google Maps, city hall open data portals, and public shapefiles) and computing all distances with the \texttt{geopy} library. Rather than using point approximations, we represented streets as polylines and calculated geodesic distances from each property to the nearest segment, yielding more realistic proximity measures. We also assigned each listing to the closest administrative neighborhood based on geographic centroids. To capture local market prominence, each property was also assigned a neighborhood category (\texttt{NBHD}) based on the frequency of listings in its neighborhood. This extended spatial feature set captures variability in accessibility to infrastructure and services across urban space, offering substantially higher resolution than the baseline version of the dataset and enabling more interpretable spatial dependencies in the resulting BNs.

\paragraph{Discretization of continuous variables}
To facilitate BN modeling, all continuous variables were discretized using quantile-based binning. While BNs can accommodate continuous inputs, structure learning in that setting is computationally demanding and less interpretable. Discretization enables the derivation of conditional probability tables that clearly express dependencies among variables. Following established practices \citep[e.g.,][]{beuzen2018,nojavan2017}, variables were binned into interpretable categories (e.g., \textit{Small, Medium, Large, Luxury} for area; \textit{Very Near} to \textit{Far} for distance-based features). The price per square meter variable was binned into six equally populated categories (\textit{Very Low} to \textit{Luxury}) to preserve resolution in downstream analyses. The final set of variables used for each city-specific network, including their groupings, definitions, and levels, is summarized in Table~\ref{tab:variables}.

\subsection{Discrete Bayesian Networks}

\subsubsection{Basic Principles of Bayesian Networks}

A Bayesian network (BN) is a probabilistic graphical model that represents the joint distribution of a set of random variables through a directed acyclic graph (DAG) \citep[e.g.][]{koller2009,pearl2014}. Each node corresponds to a variable, and each directed edge encodes a direct dependency. The absence of an edge implies conditional independence, formalized through the d-separation criterion \citep{pearl2014}.

Given a set of $n$ discrete variables $X_1, \dots, X_n$, a BN defines the joint probability distribution as a product of conditional probabilities:

\begin{equation}
P(X_1, X_2, \dots, X_n) = \prod_{i=1}^n P(X_i \mid \text{Parents}(X_i)),
\end{equation}

\noindent
where $\text{Parents}(X_i)$ denotes the set of immediate predecessors of $X_i$ in the DAG. This factorization allows BNs to represent high-dimensional distributions in a compact and interpretable form, especially when many conditional independencies are present. As a result, BNs are widely used in fields where uncertainty modeling and transparent reasoning are essential.

\subsubsection{Learning Bayesian Networks}

Although BNs can be specified manually using expert knowledge \citep{barons2022balancing,nyberg2022bard}, in this study we adopt a fully data-driven approach. Both the structure of the network (i.e., the DAG) and the associated conditional probabilities are learned from data.

We implement a score-based structure learning strategy using the Tabu search algorithm \citep{tsamardinos2006max} guided by the Bayesian Information Criterion (BIC), a widely used metric that balances model fit and complexity. Given a graph structure $G$ and dataset $D$, the BIC score is defined as:

\begin{equation}
\text{Score}_{\text{BIC}}(G \mid D) = \log P(D \mid G, \hat{\theta}) - \frac{d}{2} \log N,
\end{equation}

\noindent where $P(D \mid G, \hat{\theta})$ is the likelihood of the data under the maximum likelihood parameters $\hat{\theta}$, $d$ is the number of free parameters in the model, and $N$ is the sample size. For discrete variables, the log-likelihood can be expressed as:

\begin{equation}
\log P(D \mid G, \hat{\theta}) = \sum_{i=1}^{n} \sum_{j=1}^{q_i} \sum_{k=1}^{r_i} N_{ijk} \log \frac{N_{ijk}}{N_{ij}},
\end{equation}

\noindent where $X_i$ is a node with $r_i$ possible states and $q_i$ parent configurations, and $N_{ijk}$ is the count of observations with $X_i = x_k$ and parents in configuration $x_j$.

To improve robustness and mitigate overfitting, we apply a non-parametric bootstrap procedure \citep{scutari2013}. The structure learning process is repeated over 2000 bootstrap samples. A consensus network is then constructed by retaining only the edges that appear in at least 50\% of the learned structures. To enhance interpretability, we constrain the target variable \texttt{PRICE} (representing binned price per square meter) to have no outgoing edges during structure learning. Any cycles detected in the consensus graph are resolved by iteratively removing the weakest arc in each cycle.

Finally, the conditional probability tables were estimated using Bayesian parameter learning with a Dirichlet prior and an equivalent sample size of 1. This regularization approach ensures that all probability entries remain strictly positive, avoiding zero estimates in sparse configurations. Given observed counts $N_{ijk}$ for variable $X_i$ in state $x_k$ and parent configuration $x_j$, the estimate of the conditional probability is given by:

\begin{equation}
\hat{P}(X_i = x_k \mid \text{Pa}(X_i) = x_j) = \frac{N_{ijk} + \alpha}{N_{ij} + r_i \alpha},
\end{equation}

\noindent where $N_{ij} = \sum_k N_{ijk}$, $r_i$ is the number of states of $X_i$, and $\alpha = 1$ corresponds to a uniform Dirichlet prior. This estimator smoothly shrinks all probabilities away from 0 and is robust to low-frequency patterns in the data \citep{heckerman1995learning}.

All networks were learned using the \texttt{bnlearn} package in R \citep{scutari2010learning} and the resulting models are freely available within the \texttt{bnRep} package \citep{leonelli2025bnrep}.

\subsubsection{Analyses Based on the Bayesian Network Model}

Having constructed the  BN, we analyze its structure and implications using a comprehensive suite of inference and sensitivity techniques. These methods provide insight into the model’s internal logic, highlight key relationships among variables, and quantify the relevance and robustness of features influencing residential property prices. Our approach combines well-established tools with recent methodological advances.

To understand how different property configurations relate to price outcomes, we apply three complementary inference procedures. First, we compute the most probable explanation (MPE) for each state of the \texttt{PRICE} variable. The MPE identifies the most likely joint configuration of all other variables, conditional on observing a given price level, offering interpretable “typical scenarios” for each market segment \citep{kwisthout2011most}. Second, we perform evidence propagation: each variable is fixed to a specific state, and the resulting change in the marginal distribution of \texttt{PRICE} is recorded. This enables comparison of directional effects and highlights variables with the strongest influence on pricing. Third, we conduct scenario-based inference by fixing multiple variables simultaneously to simulate realistic housing profiles, such as centrally located luxury units or compact suburban dwellings, and examining the resulting price distributions.

We then assess the marginal importance of each variable using three complementary measures. First, mutual information quantifies the overall statistical dependence between each feature and \texttt{PRICE}, capturing both linear and non-linear associations \citep[computed with the \texttt{bnmonitor} R package][]{leonelli2023sensitivity}. Second, we use variance component Sobol indices to estimate how much the observation of each variable reduces the uncertainty in price categories, thereby isolating its individual contribution to overall variability in the target \citep{ballester2022computing}. Finally, we compute diameter-based arc strength, which measures the maximum influence a parent variable can exert on its child via the total variation distance across the conditional probability table \citep{leonelli2025diameter}.

We also investigate local parameter sensitivity to understand how small changes in individual parameters affect the probability of price outcomes. To visualize this, we use tornado plots, which display the most impactful entries in the conditional probability tables for a given price level. Each sensitivity value quantifies how much the posterior probability can shift in response to perturbations of a single conditional probability entry \citep{ballester2023yodo}.

Taken together, these analyses provide a layered understanding of the BN’s probabilistic structure. They enable us to identify not only which structural and spatial features most strongly influence price formation, but also how robust these conclusions are to uncertainty in model parameters. The interpretability and flexibility of BNs make them particularly well suited to support urban policy, price scenario analysis, and transparent decision-making in real estate markets.

\section{Results}

\begin{figure}
\centering
\includegraphics[width=\textwidth]{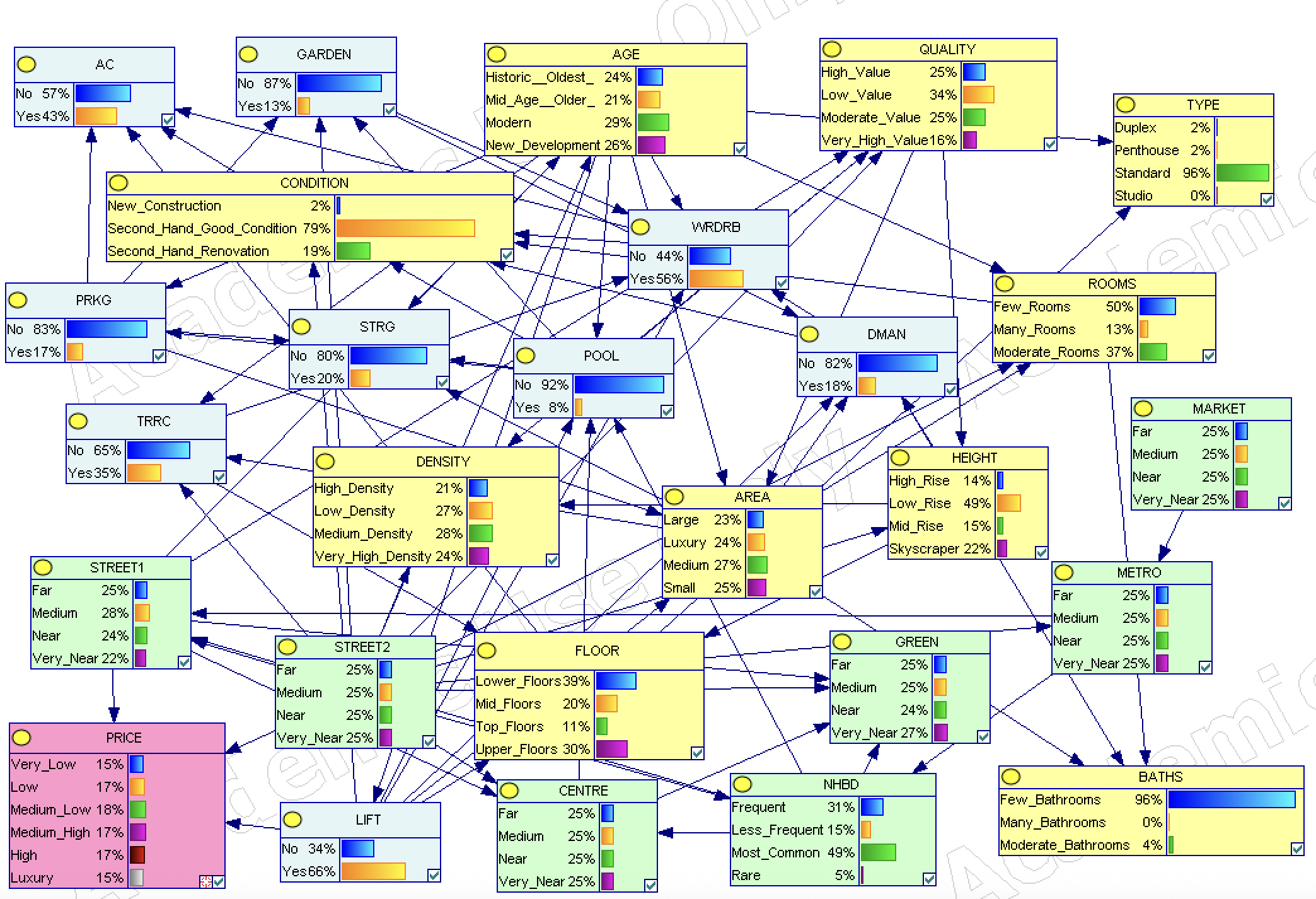}
\caption{Learned Bayesian network for Madrid. Nodes are colored by variable group: structural features (yellow), spatial indicators (green), amenities (blue), and the target variable \texttt{PRICE} (pink). Visualization produced using GeNIe.}
\label{fig:bn_madrid}
\end{figure}

\subsection{Structure and interpretation of the learned Bayesian networks}

\begin{figure}
\centering
\includegraphics[width=\textwidth]{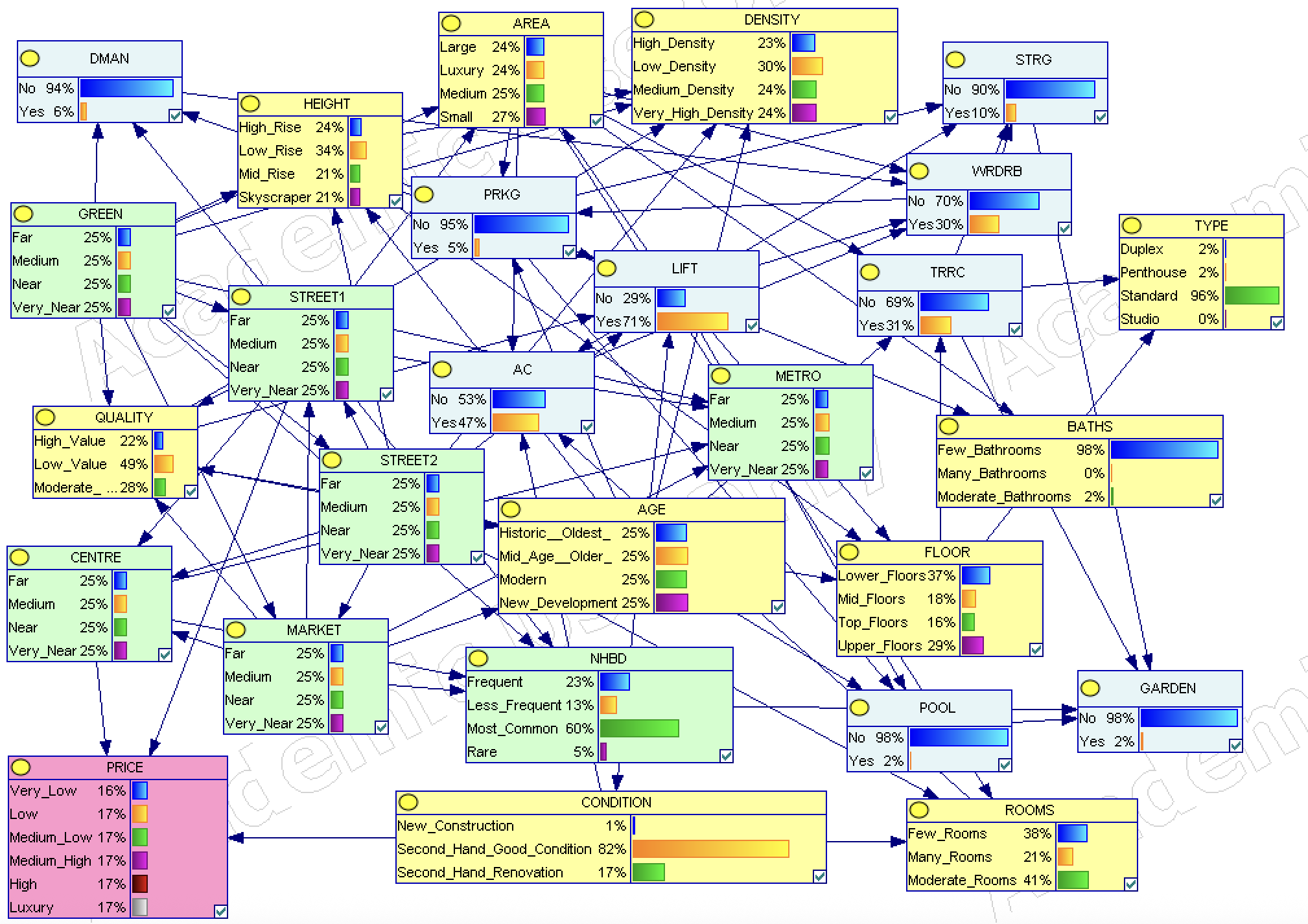}
\caption{Learned Bayesian network for Barcelona. Nodes are colored by variable group: structural features (yellow), spatial indicators (green), amenities (blue), and the target variable \texttt{PRICE} (pink). Visualization produced using GeNIe.}

\label{fig:bn_barcelona}
\end{figure}

\begin{figure}
\centering
\includegraphics[width=\textwidth]{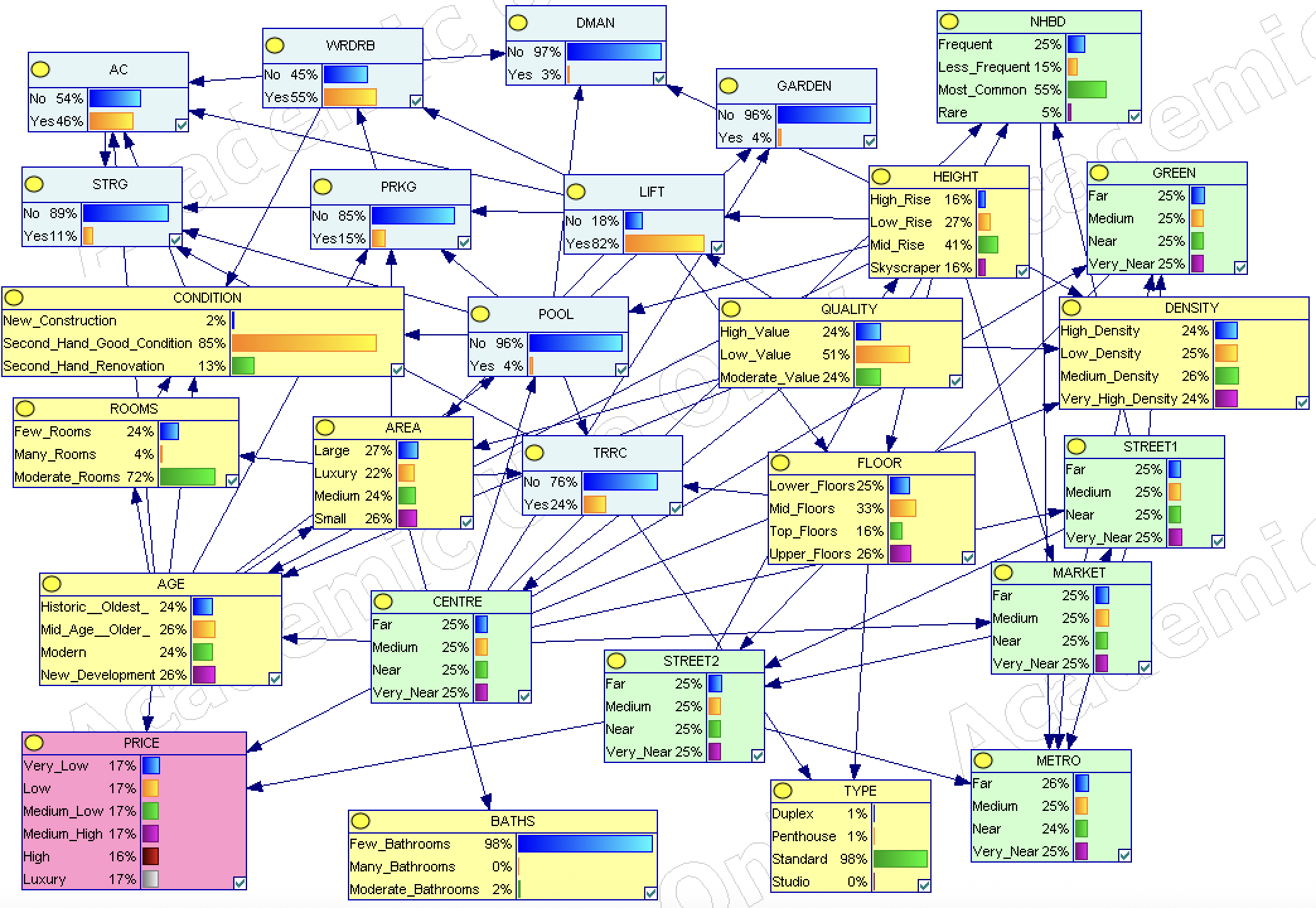}
\caption{Learned Bayesian network for Valencia. Nodes are colored by variable group: structural features (yellow), spatial indicators (green), amenities (blue), and the target variable \texttt{PRICE} (pink). Visualization produced using GeNIe.}
\label{fig:bn_valencia}
\end{figure}

Figures~\ref{fig:bn_madrid}, \ref{fig:bn_barcelona}, and \ref{fig:bn_valencia} show the structure of the learned BNs for Madrid, Barcelona, and Valencia. While all three graphs are defined over the same set of 27 variables, their connectivity and dependency patterns differ substantially. These contrasts provide insight into the distinct urban logics underlying price formation in each city.

\paragraph{Madrid}

The Madrid network contains 77 directed edges, with \texttt{PRICE} having three direct parents: \texttt{LIFT}, \texttt{STREET1}, and \texttt{STREET2}. These represent accessibility and proximity to major corridors (La Castellana and Gran Vía) confirming their role as spatial anchors in the capital’s housing market. Structural variables like \texttt{AREA} and \texttt{ROOMS} connect only indirectly to price via spatial and amenity paths, suggesting that size and configuration affect price primarily through interactions with comfort and location. Highly central nodes include \texttt{STREET2}, \texttt{AGE}, and \texttt{POOL}, reflecting how newer or centrally located buildings tend to co-occur with high-end features. The structure captures a layered market in which accessibility, modernity, and premium amenities align to differentiate price segments.

\paragraph{Barcelona}

Barcelona’s network is the densest of the three, with 81 edges. The parents of \texttt{PRICE} are \texttt{CONDITION}, \texttt{CENTRE}, and \texttt{STREET1} (La Diagonal). Here, price is linked more to typology and centrality than to amenities. Variables such as \texttt{LIFT}, \texttt{POOL}, and \texttt{PARKING} are more peripheral, connected through intermediaries like \texttt{TYPE} and \texttt{QUALITY}. Central roles are played by \texttt{STREET2} (Las Ramblas), \texttt{AGE}, and \texttt{GREEN}, highlighting the influence of spatial form and environment. The presence of \texttt{CONDITION} as a direct parent suggests a price logic tied to property status, possibly influenced by zoning constraints and renovation activity, with dwelling classification acting as a gateway to pricing.

\paragraph{Valencia}

Valencia’s network is the sparsest, with 69 edges and a simpler pattern of dependencies. \texttt{PRICE} has three direct parents: \texttt{AGE}, \texttt{CENTRE}, and \texttt{STREET2} (Avenida Blasco Ibáñez). These point to a leaner pricing logic dominated by temporal and spatial fundamentals. Fine-grained structural and lifestyle variables such as \texttt{CONDITION}, \texttt{ROOMS}, or \texttt{LIFT} play no direct role and appear only marginally in the graph. This flatter structure likely reflects a more homogeneous housing stock, where property age and proximity to the city core are sufficient to explain price tier. The absence of strong amenity-related paths may indicate limited premium differentiation in Valencia’s real estate market.

\paragraph{Comparative observations}

Across all three cities, spatial variables,especially \texttt{CENTRE}, \texttt{STREET1}, and \texttt{STREET2}, play a central role in price formation. However, their placement and influence vary: \texttt{STREET2} is a direct parent of \texttt{PRICE} in Madrid and Valencia but serves more as a mediator in Barcelona. Similarly, \texttt{LIFT} is a direct input only in Madrid, while \texttt{CONDITION} appears in that role only in Barcelona. Valencia, in contrast, leans more on coarse but informative variables like \texttt{AGE}, with fewer overall links. Amenities are generally peripheral, except in Madrid, where nodes like \texttt{POOL} and \texttt{WRDRB} are more tightly integrated. In Barcelona and Valencia, amenities play a secondary role, consistent with a pricing logic driven more by typology and spatial fundamentals, respectively.

Overall, the networks reveal how different urban dynamics shape the structure of price formation. In Madrid, accessibility and amenity coalesce into a layered segmentation; in Barcelona, typology and centrality dominate; in Valencia, price is determined largely by age and location. These distinctions underscore the value of BNs in uncovering interpretable, city-specific patterns in housing markets.

\subsection{Probabilistic Inference and Key Drivers}

\subsubsection{Most Probable Explanations for Price Levels}

\begin{table}
\centering
\footnotesize
\scalebox{0.60}{
\begin{tabular}{l
>{\columncolor{gray!2}}c  % No Evidence
>{\columncolor{gray!5}}c
>{\columncolor{gray!10}}c
>{\columncolor{gray!15}}c
>{\columncolor{gray!20}}c
>{\columncolor{gray!25}}c
>{\columncolor{gray!30}}c}
\toprule
\multicolumn{8}{c}{\textbf{Barcelona}} \\
\midrule
\textbf{Variable} & \textbf{No Evidence} & \textbf{Very Low} & \textbf{Low} & \textbf{Medium Low} & \textbf{Medium High} & \textbf{High} & \textbf{Luxury} \\
\midrule
AC       & No  & No  & No  & No  & No  & No  & No \\
AGE      & New Development & Modern & New Development & Modern & Modern & Mid Age & Mid Age \\
AREA     & Large & Large & Small & Small & Small & Large & Large \\
CENTRE   & Near & Near & Medium & Far & Far & Near & Near \\
DENSITY  & Very High & Very High & Very High & Low & Low & Very High & Very High \\
FLOOR    & Lower & Lower & Lower & Lower & Lower & Lower & Lower \\
GREEN    & Far & Far & Very Near & Very Near & Near & Far & Far \\
HEIGHT   & High Rise & Skyscraper & High Rise & Low Rise & Low Rise & High Rise & High Rise \\
LIFT     & Yes & Yes & Yes & No & No & Yes & Yes \\
MARKET   & Medium & Medium & Far & Far & Far & Medium & Medium \\
METRO    & Very Near & Very Near & Far & Far & Very Near & Very Near & Very Near \\
NHBD     & Most Common & Most Common & Less Frequent & Frequent & Frequent & Most Common & Most Common \\
QUALITY  & Low & Low & Low & Moderate & Moderate & Low & Low \\
ROOMS    & Moderate & Moderate & Few & Few & Few & Moderate & Moderate \\
STREET1  & Very Near & Very Near & Medium & Far & Far & Very Near & Very Near \\
STREET2  & Near & Near & Medium & Far & Far & Near & Near \\
WRDRB    & No & No & No & No & No & No & No \\
\midrule
\multicolumn{8}{c}{\textbf{Madrid}} \\
\midrule
\textbf{Variable} & \textbf{No Evidence} & \textbf{Very Low} & \textbf{Low} & \textbf{Medium Low} & \textbf{Medium High} & \textbf{High} & \textbf{Luxury} \\
\midrule
AC       & No  & No  & No  & No  & Yes & Yes & Yes \\
AGE      & Historic & Mid Age & Mid Age & New Development & Historic & Historic & Historic \\
AREA     & Small & Small & Small & Small & Small & Small & Small \\

CENTRE   & Very Near & Medium & Medium & Far & Very Near & Very Near & Very Near \\
DENSITY  & Medium & Low & Low & Low & Medium & Medium & Medium \\
FLOOR    & Lower & Lower & Lower & Lower & Lower & Lower & Lower \\
GREEN    & Very Near & Medium & Medium & Far & Very Near & Very Near & Very Near \\
HEIGHT   & Low Rise & Low Rise & Low Rise & Low Rise & Low Rise & Low Rise & Low Rise \\
LIFT     & Yes & No & No & Yes & Yes & Yes & Yes \\
MARKET   & Very Near & Far & Far & Medium & Very Near & Very Near & Very Near \\
METRO    & Very Near & Far & Far & Medium & Very Near & Very Near & Very Near \\
NHBD     & Most Common & Most Common & Most Common & Most Common & Most Common & Most Common & Most Common \\
QUALITY  & Low & Very High & High & Very High & Low & Low & Low \\
ROOMS    & Few & Few & Few & Few & Few & Few & Few \\
STREET1  & Very Near & Medium & Medium & Far & Very Near & Very Near & Very Near \\
STREET2  & Very Near & Medium & Medium & Far & Very Near & Very Near & Very Near \\
WRDRB    & No & No & No & Yes & Yes & Yes & Yes \\
\midrule
\multicolumn{8}{c}{\textbf{Valencia}} \\
\midrule
\textbf{Variable} & \textbf{No Evidence} & \textbf{Very Low} & \textbf{Low} & \textbf{Medium Low} & \textbf{Medium High} & \textbf{High} & \textbf{Luxury} \\
\midrule
AC       & Yes & No & Yes & Yes & Yes & Yes & Yes \\
AGE      & Modern & Mid Age & Modern & Modern & Modern & Historic & Historic \\
AREA     & Large & Small & Large & Large & Large & Luxury & Luxury \\

CENTRE   & Far & Far & Near & Near & Near & Very Near & Very Near \\
DENSITY  & Very High & High & Medium & Medium & Medium & Medium & Medium \\
FLOOR    & Mid & Mid & Mid & Mid & Mid & Mid & Mid \\
GREEN    & Medium & Medium & Far & Far & Very Near & Medium & Very Near \\
HEIGHT   & Mid Rise & Mid Rise & Mid Rise & Mid Rise & Mid Rise & Mid Rise & Mid Rise \\
LIFT     & Yes & Yes & Yes & Yes & Yes & Yes & Yes \\
MARKET   & Very Near & Very Near & Far & Far & Near & Far & Very Near \\
METRO    & Far & Far & Very Near & Medium & Far & Very Near & Medium \\
NHBD     & Most Common & Most Common & Most Common & Most Common & Frequent & Most Common & Less Frequent \\
QUALITY  & Low & High & Low & Low & Low & Low & Low \\
ROOMS    & Moderate & Moderate & Moderate & Moderate & Moderate & Moderate & Moderate \\
STREET1  & Far & Far & Near & Near & Medium & Very Near & Near \\
STREET2  & Near & Near & Far & Medium & Very Near & Medium & Very Near \\
WRDRB    & Yes & No & Yes & Yes & Yes & Yes & Yes \\
\bottomrule
\end{tabular}}
\caption{Most probable explanations (MPE) for each price level in Barcelona, Madrid, and Valencia, including the unconditional MPE (No Evidence).}
\label{tab:mpe_all_cities_compressed}
\end{table}

We begin our analysis by examining the most probable explanation associated with each level of the \texttt{PRICE} variable. Table~\ref{tab:mpe_all_cities_compressed} summarizes, for each city, the most likely joint configuration of explanatory variables under each price category. For clarity, variables that remained constant across all price levels and cities, such as \texttt{POOL}, \texttt{TYPE}, and \texttt{CONDITION}, were omitted.

Across all cities, we observe consistent shifts in spatial proximity and building age as price increases. Higher price levels tend to be associated with properties located closer to the city center and to key corridors such as \texttt{STREET1} and \texttt{STREET2}. In both Madrid and Valencia, these variables evolve monotonically from \textit{Far} or \textit{Medium} in the lower tiers to \textit{Very Near} in the highest, confirming spatial accessibility as a key driver of value. In Barcelona, proximity to Diagonal and Las Ramblas is already present in the lowest price bands, reflecting the city’s denser urban fabric and greater baseline accessibility. Differences also emerge in the role of spatial amenities such as metro access and green space: in Barcelona, very low and low price bands are more likely to occur near green areas and far from the metro, whereas in Madrid and Valencia, metro proximity is consistently linked to higher price tiers.

Structural contrasts are also evident. In Barcelona, the highest price categories are linked to newer buildings and larger units, typically in modern or mid-aged developments with high-rise construction. In contrast, Madrid’s luxury segment is more frequently found in historic buildings with smaller surface area, suggesting a premium for classical architecture and location over internal size. Valencia exhibits a more transitional pattern, where luxury units tend to be newer and larger but show little variation in internal configuration, for instance the number of rooms remains \textit{Moderate} across all tiers.

\subsubsection{Price Sensitivity Under Evidence Propagation}

\begin{table}
\centering
\footnotesize
\scalebox{0.70}{
\begin{tabular}{l
>{\columncolor{gray!5}}c
>{\columncolor{gray!10}}c
>{\columncolor{gray!15}}c
>{\columncolor{gray!20}}c
>{\columncolor{gray!25}}c
>{\columncolor{gray!30}}c}
\toprule
\multicolumn{7}{c}{\textbf{Barcelona}} \\
\midrule
\textbf{Variable (Value)} & \textbf{Very Low} & \textbf{Low} & \textbf{Medium Low} & \textbf{Medium High} & \textbf{High} & \textbf{Luxury} \\
\midrule
CENTRE = Far & 0.0856 & 0.2536 & 0.1441 & \textbf{0.3970} & 0.0613 & \textit{0.0583} \\
CENTRE = Near & 0.2098 & 0.1152 & 0.1725 & \textit{0.0768} & \textbf{0.2242} & 0.2015 \\
CONDITION = New Construction & 0.2450 & \textit{0.0249} & 0.0889 & 0.0188 & \textbf{0.3664} & 0.2560 \\
NHBD = Less Frequent & 0.1113 & \textbf{0.2369} & 0.1611 & 0.3163 & 0.0882 & \textit{0.0862} \\
NHBD = Rare & 0.1086 & 0.2338 & 0.1591 & \textbf{0.3287} & 0.0880 & \textit{0.0820} \\
STREET1 = Far & 0.0859 & 0.2561 & 0.1477 & \textbf{0.4216} & 0.0588 & \textit{0.0300} \\
STREET1 = Near & 0.1947 & 0.1452 & 0.1752 & \textit{0.0550} & 0.2055 & \textbf{0.2245} \\
STREET1 = Very Near & 0.1918 & 0.0769 & 0.1433 & \textit{0.0310} & 0.2572 & \textbf{0.2998} \\
STREET2 = Far & 0.0879 & 0.2525 & 0.1461 & \textbf{0.3891} & 0.0636 &\textit{0.0607} \\
STREET2 = Near & 0.2093 & 0.0909 & 0.1522 & \textit{0.0506} & 0.2464 & \textbf{0.2505} \\
\midrule
\multicolumn{7}{c}{\textbf{Madrid}} \\
\midrule
\textbf{Variable (Value)} & \textbf{Very Low} & \textbf{Low} & \textbf{Medium Low} & \textbf{Medium High} & \textbf{High} & \textbf{Luxury} \\
\midrule
AGE = Historic  & \textit{0.0400} & 0.0592 & 0.1195 & 0.2206 & 0.2657 & \textbf{0.2951} \\
CENTRE = Far & \textbf{0.2911} & 0.2709 & 0.2243 & 0.1218 & 0.0687 & \textit{0.0231} \\
CENTRE = Medium & 0.2265 & \textbf{0.2602} & 0.2382 & 0.1315 & 0.0861 & \textit{0.0575} \\
CENTRE = Very Near & \textit{0.0106} & 0.0239 & 0.0878 & 0.2209 & 0.2960 & \textbf{0.3609} \\
STREET1 = Far & 0.3053 & \textbf{0.3362} & 0.2302 & 0.0798 & 0.0373 & \textit{0.0113} \\
STREET1 = Near & \textit{0.0548} & 0.0759 & 0.1438 & 0.2154 & \textbf{0.2621} & 0.2480 \\
STREET1 = Very Near & \textit{0.0221} & 0.0496 & 0.1255 & 0.2347 & 0.2557 & \textbf{0.3124} \\
STREET2 = Far & \textbf{0.2988} & 0.2701 & 0.2222 & 0.1213 & 0.0675 & \textit{0.0202} \\
STREET2 = Medium & 0.2324 & \textbf{0.2702}& 0.2448 & 0.1227 & 0.0771 & \textit{0.0527} \\
STREET2 = Very Near & \textit{0.0056} & 0.0161 & 0.0793 & 0.2172 & 0.3045 & \textbf{0.3773} \\
\midrule
\multicolumn{7}{c}{\textbf{Valencia}} \\
\midrule
\textbf{Variable (Value)} & \textbf{Very Low} & \textbf{Low} & \textbf{Medium Low} & \textbf{Medium High} & \textbf{High} & \textbf{Luxury} \\
\midrule
AGE = New Development & \textit{0.0395} & 0.1066 & 0.1652 & 0.2098 & 0.2269 & \textbf{0.2520} \\
CENTRE = Far & \textbf{0.3549} & 0.2100 & 0.1288 & 0.1262 & 0.0998 & \textit{0.0803} \\
CENTRE = Medium & 0.2125 & 0.2230 & \textbf{0.2292} & 0.1659 & 0.1072 & \textit{0.0623} \\
CENTRE = Very Near & \textit{0.0107} & 0.0372 & 0.0727 & 0.1569 & 0.2861 & \textbf{0.4364} \\
GREEN = Far & \textbf{0.2373} & 0.2345 & 0.2099 & 0.1678 & 0.0981 & \textit{0.0523} \\
QUALITY = High Value & \textbf{0.3070} & 0.2138 & 0.1668 & 0.1336 & 0.0987 & \textit{0.0801} \\
STREET1 = Far & \textbf{0.3441} & 0.2207 & 0.1429 & 0.1263 & 0.0947 & \textit{0.0712} \\
STREET1 = Very Near & \textit{0.0411} & 0.0847 & 0.1287 & 0.1805 & 0.2482 & \textbf{0.3169} \\
STREET2 = Far & \textbf{0.2843} & 0.2570 & 0.2124 & 0.1501 & 0.0550 & \textit{0.0411} \\
STREET2 = Very Near & \textit{0.0427} & 0.0964 & 0.1554 & 0.1719 & 0.2122 & \textbf{0.3214} \\
\bottomrule
\end{tabular}
}
\caption{Posterior probabilities for each price level under the top 10 variable-value pairs (by symmetrized KL divergence) for Barcelona, Madrid, and Valencia. Columns are ordered from lowest to highest price level. For each row, the highest probability is shown in \textbf{bold} and the lowest in \textit{italics}.}
\label{tab:price_conditional_top10}
\end{table}

Table~\ref{tab:price_conditional_top10} presents the conditional price distributions under the ten most influential variable-value pairs in each city, based on symmetrized KL divergence from the marginal. For each row, the highest posterior probability is in bold and the lowest in italics. The patterns confirm that spatial accessibility is the dominant driver of high price segments: proximity to the city center and key corridors (\texttt{CENTRE}, \texttt{STREET1}, \texttt{STREET2}) consistently shifts mass toward the upper price tiers.

Across cities, secondary drivers differ. In Madrid, luxury pricing is associated with historic buildings and strong centrality effects. In Valencia, by contrast, newer developments dominate the high-price profiles, with age and accessibility acting as structural fundamentals. Barcelona exhibits a more bimodal structure: properties in rare neighborhoods or with \texttt{CONDITION = New Construction} are overrepresented at both the lowest and highest price levels, pointing to a polarized market shaped by typology and classification.

These results reinforce the spatial and structural patterns seen in the learned networks, and further illustrate how distinct urban logics shape price expectations across cities.

\begin{table}
\centering
\footnotesize
\scalebox{0.75}{
\begin{tabular}{p{4cm} p{13cm}}
\toprule
\textbf{Scenario} & \textbf{Evidence (Variable = Value)} \\
\midrule
\textbf{Luxury Core} & CENTRE = Very Near, STREET1 = Very Near, QUALITY = High Value, AREA = Luxury, LIFT = Yes \\
\textbf{Green Comfort} & AREA = Large, GREEN = Very Near, ROOMS = Many Rooms, TRRC = Yes, POOL = Yes \\
\textbf{Urban Compact} & AREA = Small, AGE = Historic, DENSITY = Very High, CENTRE = Near, MARKET = Very Near \\
\textbf{Metro Suburb} & CENTRE = Far, METRO = Very Near, HEIGHT = Mid Rise, CONDITION = Second Hand Good Condition \\
\textbf{Modern Convenience} & AGE = Modern, LIFT = Yes, MARKET = Near, METRO = Near \\
\textbf{Peripheral Standard} & CENTRE = Far, STREET1 = Far, STREET2 = Far \\
\textbf{Green Retreat} & GREEN = Very Near, STREET1 = Near, CONDITION = Second Hand Good Condition \\
\bottomrule
\end{tabular}}
\caption{Scenario definitions used for multi-variable inference. Each scenario corresponds to a distinct configuration of structural and spatial features that jointly influence predicted price levels.}
\label{tab:scenario_table}
\end{table}

\subsubsection{Scenario-Based Price Distributions}

\begin{figure}
\centering
\includegraphics[width=\textwidth]{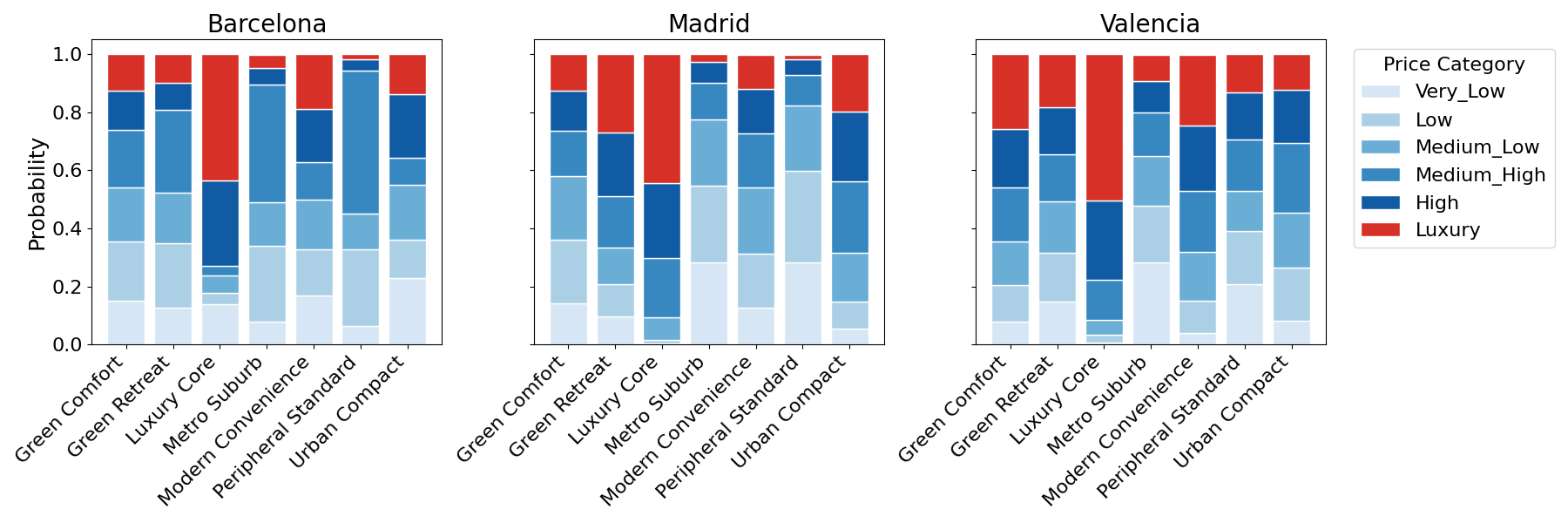}
\caption{Posterior distribution of price categories under each scenario across the three cities. Bars are stacked by price tier and grouped by scenario (left to right: Barcelona, Madrid, Valencia). See Table~\ref{tab:scenario_table} for scenario definitions.}
\label{fig:scenario_plot}
\end{figure}

BNs allow for joint inference under arbitrary configurations of variables, enabling the simulation of realistic housing profiles. Table~\ref{tab:scenario_table} defines seven such scenarios, covering a range of structural and spatial settings, from luxury dwellings in premium districts to compact or suburban units. Figure~\ref{fig:scenario_plot} displays the resulting posterior distributions over price categories in each city.

As expected, the \textit{Luxury Core} scenario shifts mass sharply toward the highest price tiers in all three cities, particularly in Valencia, where more than 50\% of listings fall into the \textit{Luxury} category. By contrast, the \textit{Urban Compact} and \textit{Peripheral Standard} scenarios concentrate probability in the lower half of the distribution, especially in Madrid. The \textit{Green Comfort} and \textit{Green Retreat} scenarios produce broader distributions, reflecting trade-offs between area, amenities, and location.

City-specific patterns further emerge. In Madrid, centrality effects dominate: the \textit{Metro Suburb} scenario shows a marked drop in luxury pricing. In Barcelona, the \textit{Modern Convenience} and \textit{Green Comfort} scenarios sustain moderate probabilities in the upper tiers, consistent with the city’s dense structure and mixed amenity presence. Valencia exhibits flatter transitions across scenarios, although the \textit{Luxury Core} and \textit{Green Comfort} configurations still yield clear upward shifts in expected price.

These results demonstrate the versatility of BNs in scenario-based housing analysis, enabling nuanced probabilistic forecasts under complex but interpretable feature combinations.

\subsection{Sensitivity analysis}

\subsubsection{Arc Robustness}

The diameter-based arc strengths confirm that the structures learned are highly robust. Most retained edges show values well above 0.5, with key dependencies (such as \texttt{STREET1} $\rightarrow$ \texttt{PRICE} in Madrid and Barcelona, and \texttt{CENTRE} $\rightarrow$ \texttt{PRICE} in Valencia) exceeding 0.85. Spatial pathways generally exhibit the strongest connections, consistent with their centrality in the inferred networks. Structural links such as \texttt{AGE} $\rightarrow$ \texttt{PRICE} in Valencia and \texttt{CONDITION} $\rightarrow$ \texttt{PRICE} in Barcelona also display high arc-strength values, reinforcing their role in shaping price tiers. For brevity, we do not report the full list of arc strengths, but note that the overall pattern confirms the stability and interpretability of the learned structures.

\subsubsection{Marginal Sensitivity}

\begin{table}
\centering
\footnotesize
\renewcommand{\arraystretch}{1.2}
\scalebox{0.8}{
\begin{tabular}{lcccccc}
\toprule
 & \multicolumn{2}{c}{\textbf{Barcelona}} & \multicolumn{2}{c}{\textbf{Madrid}} & \multicolumn{2}{c}{\textbf{Valencia}} \\
\cmidrule(lr){2-3} \cmidrule(lr){4-5} \cmidrule(lr){6-7}
\textbf{Variable} & MI & Sobol & MI & Sobol & MI & Sobol \\
\midrule
CENTRE    & 15.83 (1) & 18.03 (3)  & 18.76 (3) & 60.25 (2)  & 17.44 (1) & 55.52 (1) \\
STREET2   & 15.31 (2) & 20.75 (2)  & 21.24 (1) & 66.73 (1)  & 9.45 (2)  & 28.69 (3) \\
STREET1   & 14.29 (3) & 60.85 (1)  & 18.82 (2) & 60.04 (3)  & 9.31 (3)  & 32.88 (2) \\
NHBD      & 9.95 (4)  & 14.12 (4)  & 2.52 (8)  & 9.15 (8)   & 1.29 (8)  & 3.34 (9) \\
MARKET    & 4.05 (5)  & 5.83 (7)   & 0.14 (19) & 0.37 (20)  & 0.70 (10) & 2.62 (10) \\
CONDITION & 2.17 (6)  & 2.46 (9)  & 0.02 (24) & 0.02 (25)  & 0.06 (19) & 0.15 (20) \\
AGE       & 2.04 (7)  & 0.60 (15)  & 6.62 (4) & 22.81 (4)  & 8.10 (4) & 26.76 (4) \\
QUALITY   & 1.85 (8)  & 11.50 (5)  & 5.77 (5)  & 20.29 (5)  & 4.26 (5)  & 15.99 (5) \\
AREA      & 0.74 (9)  & 2.83 (8)  & 0.89 (12) & 3.15 (12)  & 1.12 (9) & 4.24 (8) \\
DENSITY   & 0.58 (10) & 1.50 (11)  & 1.08 (10)& 3.85 (10)  & 0.52 (13) & 1.57 (13) \\
ROOMS     & 0.40 (11) & 0.43 (17)  & 0.61 (13) & 2.13 (13)   & 0.13 (16) & 0.49 (16) \\
GREEN     & 0.33 (12) & 9.88 (6)   & 4.48 (6)  & 15.49 (6)  & 3.78 (6) & 12.10 (6) \\
HEIGHT    & 0.33 (13) & 1.25 (13)  & 2.08 (9) & 7.48 (9)  & 0.57 (12) & 1.79 (12) \\
DMAN      & 0.32 (14) & 1.41 (12)  & 0.43 (15) & 1.42 (15)  & 0.04 (22) & 0.11 (23) \\
AC        & 0.25 (15) & 0.20 (19)  & 0.01 (25) & 0.00 (26)  & 0.09 (17) & 0.34 (17) \\
LIFT      & 0.19 (16) & 1.81 (10)  & 3.91 (7)  & 12.62 (7)  & 0.65 (11) & 2.43 (11) \\
FLOOR     & 0.13 (17) & 0.52 (16)  & 0.28 (17) & 0.93 (16)  & 0.09 (18) & 0.28 (18) \\
METRO     & 0.10 (18) & 0.68 (14)  & 0.14 (20) & 0.37 (19)  & 1.99 (7) & 6.96 (7) \\
WRDRB     & 0.08 (19) & 0.22 (18)  & 0.10 (21) & 0.22 (22)  & 0.05 (20) & 0.17 (19) \\
GARDEN    & 0.03 (20) & 0.03 (23)  & 0.57 (14) & 1.77 (14)  & 0.03 (23) & 0.06 (25) \\
BATHS     & 0.03 (21) & 0.10 (20)   & 0.07 (22)  & 0.26 (21)   & 0.03 (24)  & 0.13 (22) \\
TRRC      & 0.01 (22) & 0.08 (21)  & 0.94 (11) & 3.32 (11)  & 0.02 (25) & 0.09 (24) \\
PRKG      & 0.01 (23) & 0.69 (22)  & 0.20 (18) & 0.55 (18)  & 0.50 (14) & 1.54 (14) \\
STRG      & 0.01 (23) & 0.02 (24)  & 0.05 (23) & 0.10 (23)  & 0.04 (21) & 0.14 (21) \\
POOL      & 0.00 (24) & 0.01 (26)  & 0.31 (16) & 0.57 (17)  & 0.21 (15) & 0.57 (15) \\
TYPE      & 0.00 (25) & 0.01 (25)  & 0.01 (26) & 0.03 (24)  & 0.00 (26) & 0.01 (26) \\
\bottomrule
\end{tabular}
}
\caption{Mutual information and Sobol indices (×100) for each variable with respect to price per square meter in the city-specific networks. Variables are sorted by MI ranking in Barcelona. Ranks are shown in parentheses.}
\label{tab:sobol_mi_sorted_bcn}
\end{table}

Table~\ref{tab:sobol_mi_sorted_bcn} summarizes the marginal sensitivity of each variable with respect to \texttt{PRICE}, using both mutual information and Sobol indices. Spatial accessibility dominates across all three cities, with distance to the center (\texttt{CENTRE}), the main avenue (\texttt{STREET1}: Diagonal, La Castellana, Blasco Ibáñez), and the secondary corridor (\texttt{STREET2}: Las Ramblas, Gran Vía, Reino de Valencia) consistently ranked among the top three variables. However, important contrasts emerge beyond this spatial core.

In Madrid, terrace availability (\texttt{TRRC}) stands out as notably more influential than in the other cities, reflecting the premium attached to outdoor space in a landlocked market. Relatedly, amenities such as \texttt{POOL}, \texttt{LIFT}, and \texttt{GARDEN} play a more substantial role in Madrid, while air conditioning (\texttt{AC}) and proximity to supermarkets (\texttt{MARKET}) are less relevant than elsewhere.

Barcelona shows a distinct profile: dwelling classification (\texttt{CONDITION}) and local market saturation (\texttt{NHBD}) are more important than in Madrid or Valencia, suggesting that typological and neighborhood characteristics are central to price differentiation. In contrast, the influence of building age (\texttt{AGE}) is relatively minor. Distance to supermarkets (\texttt{MARKET}) ranks much higher than in the other cities, while private vehicle amenities such as \texttt{PRKG} register the lowest contribution, consistent with the city’s denser structure and stronger public transit orientation.

In Valencia, the most influential non-spatial variables are \texttt{METRO} access and \texttt{PRKG}, both of which rank substantially higher than in Barcelona and Madrid. This suggests a stronger role for mobility and convenience in price segmentation, particularly outside the historic center. At the other end of the spectrum, \texttt{DMAN} (doorman service) is nearly irrelevant in Valencia, further reinforcing the interpretation of a flatter market structure with less premium attached to lifestyle features. 

These patterns confirm that, although some price drivers are broadly shared, many exhibit city-specific relevance reflecting differences in  urban form, housing stock, and infrastructure.

\subsubsection{Local Parameter Effects}

\begin{figure}
\centering
\includegraphics[width=0.8\textwidth]{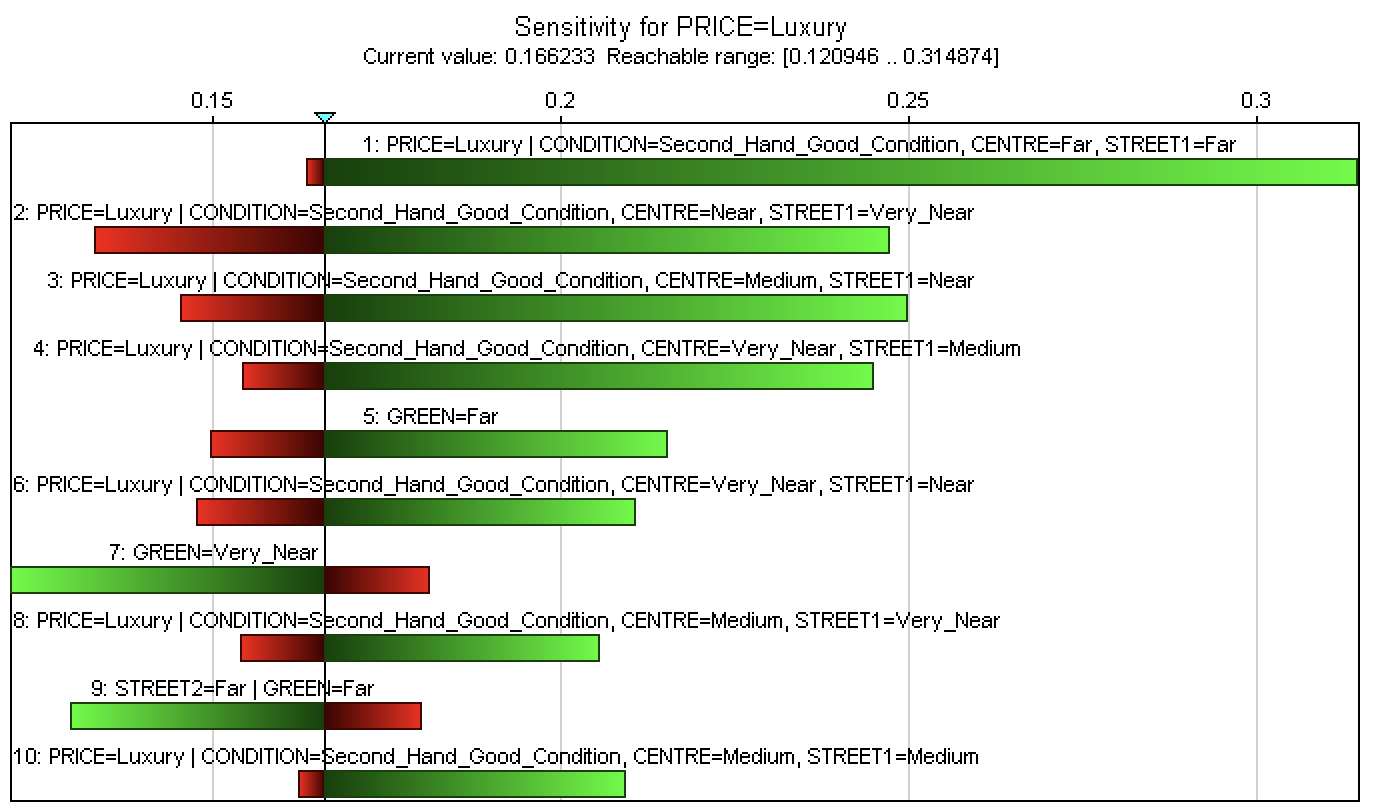}
\caption{Tornado plot for \texttt{PRICE = Luxury} in Barcelona (generated in GeNIe).}
\label{fig:tornado_barcelona}
\end{figure}

\begin{figure}
\centering
\includegraphics[width=0.8\textwidth]{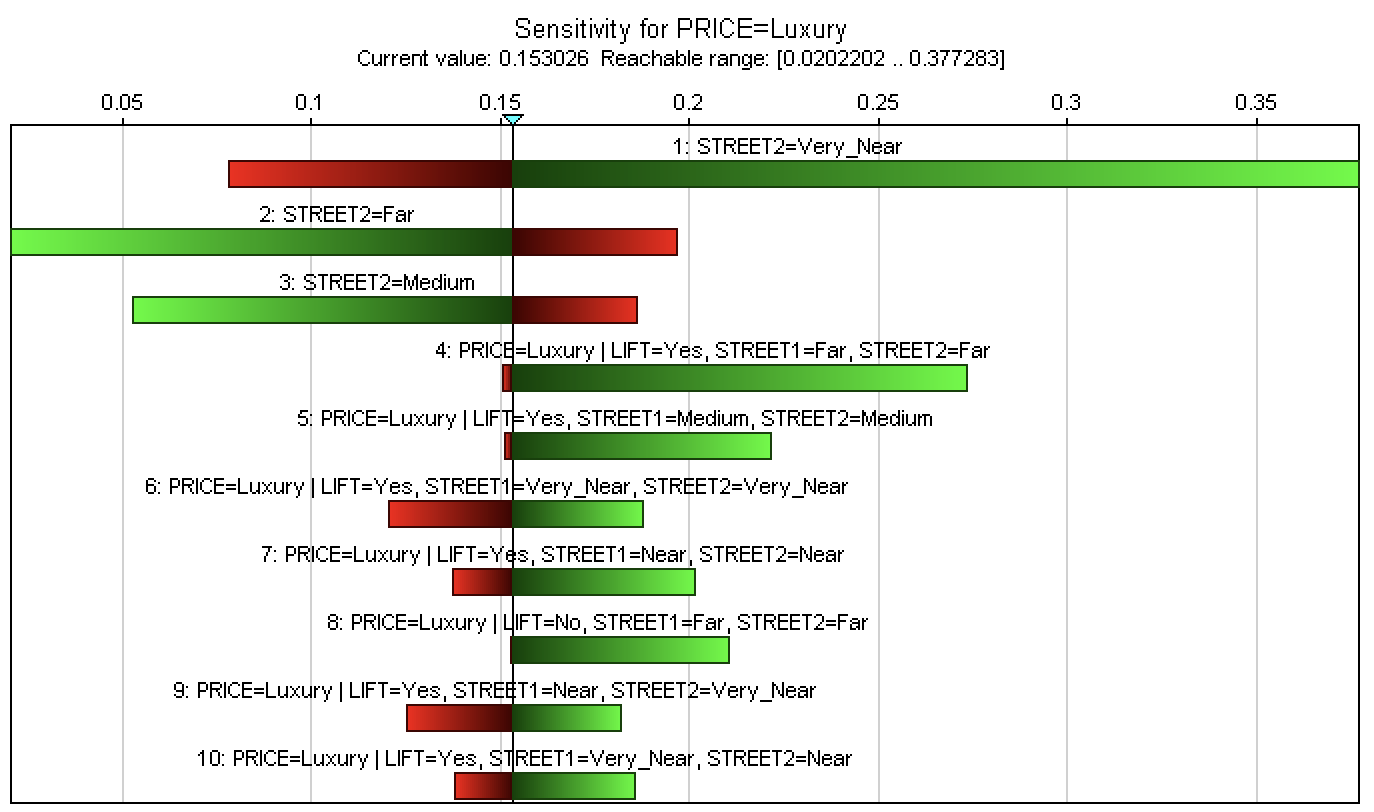}
\caption{Tornado plot for \texttt{PRICE = Luxury} in Madrid (generated in GeNIe).}
\label{fig:tornado_madrid}
\end{figure}

\begin{figure}
\centering
\includegraphics[width=0.8\textwidth]{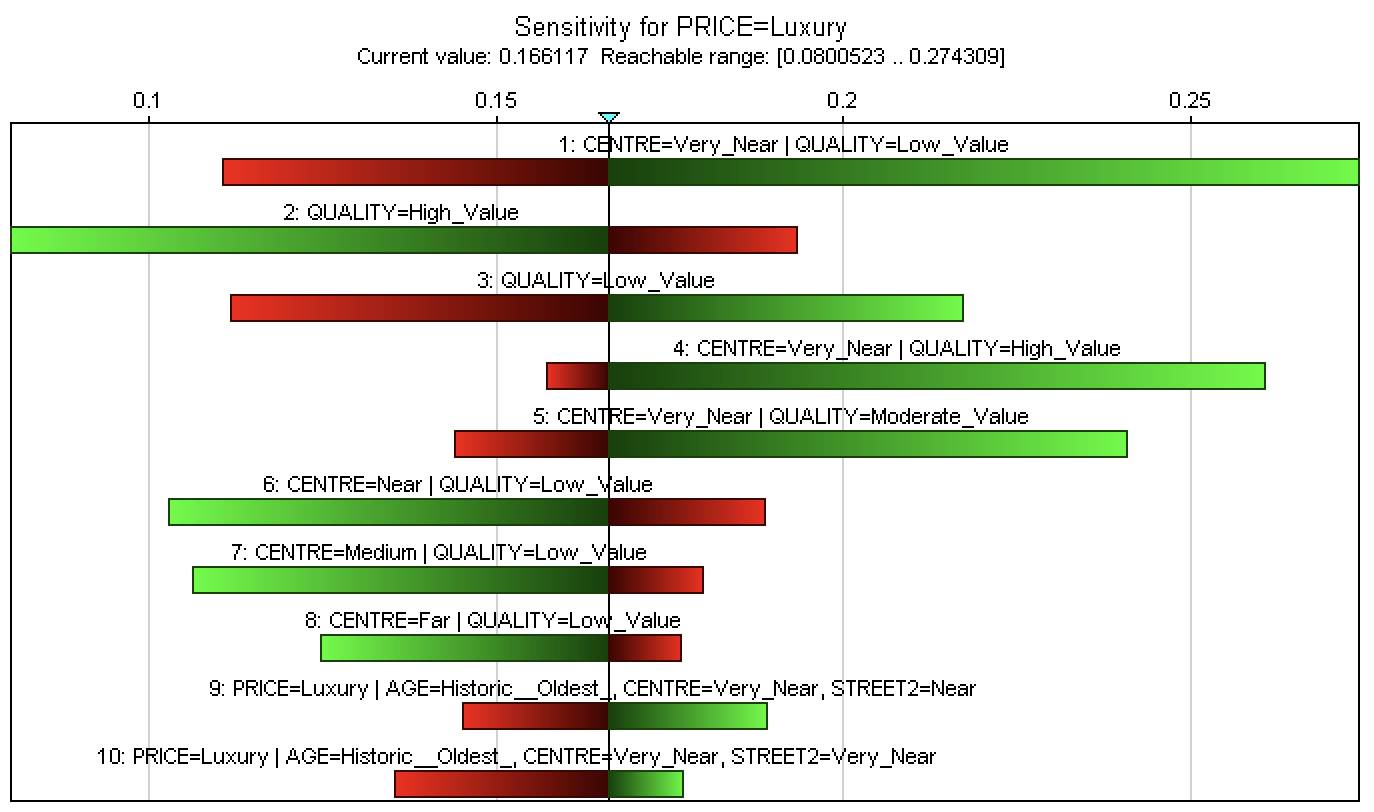}
\caption{Tornado plot for \texttt{PRICE = Luxury} in Valencia (generated in GeNIe).}
\label{fig:tornado_valencia}
\end{figure}

We assess how changes in the conditional probabilities of each variable affect the posterior distribution of \texttt{PRICE}, allowing us to identify which features exert the strongest influence on price categorization.

\ref{appendix:sensitivity_coloring} contains sensitivity-colored versions of each city-specific network, where node shading indicates overall influence on \texttt{PRICE}. Spatial variables, particularly \texttt{CENTRE}, \texttt{STREET1}, and \texttt{STREET2}, consistently show the highest sensitivity. In Barcelona, \texttt{MARKET} and \texttt{CONDITION} are also prominent; in Madrid, building quality (\texttt{QUALITY}), \texttt{LIFT}, and amenities play a stronger role. In Valencia, the emphasis falls on structural fundamentals such as \texttt{AGE}, \texttt{CENTRE}, and \texttt{GREEN}, with little impact from lifestyle features.

To illustrate these patterns more concretely, Figures~\ref{fig:tornado_barcelona}–\ref{fig:tornado_valencia} display tornado plots for the classification of \texttt{PRICE = Luxury}. These plots highlight the specific conditional probability entries whose perturbation produces the greatest shifts in predicted luxury probability.

In Barcelona, proximity to La Rambla (\texttt{STREET2 = Very Near}) is most influential, followed by Diagonal and the presence of \texttt{LIFT}, reflecting a centrality, and comfort-driven logic. In Madrid, centrality and building quality dominate, confirming a premium for centrally located, architecturally desirable units. In Valencia, the most sensitive entries involve combinations of \texttt{CONDITION}, \texttt{CENTRE}, and \texttt{STREET1}, with amenities playing a negligible role, consistent with a flatter market structure shaped by spatial and structural fundamentals.

\section{Discussion}

\subsection{Interpreting City-Specific Price Logic}

Our analysis uncovers distinct logics of price formation across Spain’s three largest housing markets. In Madrid, the networks reveal a layered structure centered on accessibility and amenities: high price tiers are anchored to corridors such as Castellana and Gran Vía, and supported by features like building quality, lift access and the presence of a terrace. This aligns with prior research on amenity clustering and corridor-driven stratification in the capital \citep{rey2023mlseg, leal2025decoding}. In Barcelona, price is more closely tied to typology and property classification. The centrality of variables like \texttt{CONDITION} and \texttt{TYPE} in the learned network supports a narrative of symbolic differentiation, consistent with the city’s dense zoning, legacy housing stock, and strong design identity. Valencia, by contrast, exhibits a flatter logic dominated by structural fundamentals such as building age and distance to the center, reflecting a market where accessibility and modernization rather than prestige or amenities drive segmentation.

These city-specific patterns resonate with the concept of functional housing submarkets, previously documented in Spain and elsewhere \citep{chasco2018scan, royuela2013housi}. Yet our findings go further: the probabilistic structure of the BNs reveals how these variables interact conditionally rather than additively, highlighting the interplay between spatial form, dwelling characteristics, and amenity value. For example, the role of green space in Barcelona differs depending on proximity to Diagonal; in Madrid, centrality modulates the importance of terraces. Such context-sensitive dependencies underscore the importance of going beyond marginal effects in housing price models, particularly in cities with complex spatial and regulatory legacies.

\subsection{Implications for Urban Policy and Market Transparency}

The interpretability and scenario-based reasoning enabled by BNs position them as particularly valuable tools for urban housing policy. In cities marked by spatial inequality, supply constraints, and rising affordability concerns (such as Madrid, Barcelona, and Valencia) planners require models that not only predict but also explain. The ability to simulate counterfactual housing profiles and inspect conditional probability shifts provides a rigorous, yet accessible, foundation for decision-making in complex, segmented markets. This is especially relevant for renovation incentives, spatially targeted subsidies, and tax valuation frameworks, where transparent price logic is essential to ensure both fairness and compliance \citep{martinez2020house}.

Moreover, the interpretability of BNs contributes directly to the legitimacy of algorithmic tools in public governance. As recent scholarship has emphasized, post hoc explanations for black-box models often fail to meet the standard of accountability required in socially impactful domains \citep{rudin2019stop}. By contrast, the probabilistic transparency of BNs allows for direct inspection of structural assumptions, variable interactions, and sensitivity outcomes, fostering both institutional trust and public engagement. This aligns with emerging demands in real estate analytics for explainable, stakeholder-facing methods that reconcile technical rigor with interpretive clarity \citep{ lorenz2023interpretable,trindade2024impacts}. In this respect, our work advances not only housing price modeling but also the broader integration of interpretable AI into the urban policy toolkit.

\subsection{Methodological Considerations and Future Research}

This study demonstrates the potential of BNs as a scalable, interpretable alternative to black-box models in housing price analytics. While BNs have been applied in domains such as environmental risk assessment and healthcare, their integration into urban real estate modeling remains limited. Our use of BNs in a multi-city, spatially structured setting, combined with data-driven learning from listing data, offers a novel methodological contribution to the field. Unlike black-box approaches that require post hoc explanation, BNs are interpretable by design: their structure encodes variable dependencies, and their conditional tables support transparent probabilistic reasoning.

That said, the approach involves important trade-offs. Discretizing continuous inputs improves interpretability and facilitates structure learning, but it may entail a loss of resolution, particularly at the extremes of the distribution. Modeling with Idealista listing data provides rich spatial coverage and detailed structural attributes, but it lacks transaction-level confirmation and may reflect listing behavior rather than final sale conditions. These limitations call for cautious interpretation and underscore the value of triangulating findings with additional data sources.

Future research may extend this framework along several dimensions. Dynamic BNs \citep{murphy2002} could capture temporal shifts in pricing logic or reveal delayed effects of infrastructural investments. Causal discovery algorithms \citep{glymour2019review} may help infer more policy-relevant dependencies, particularly in regulatory or intervention scenarios. Comparative analysis across a broader set of cities, or across time periods, could deepen understanding of how urban form, policy, and market dynamics interact. More broadly, our approach aligns with growing interest in explainable, stakeholder-facing tools for real estate and urban planning tools that not only predict but also justify, simulate, and support equitable decisions.

\section{Conclusion}

This paper has introduced a transparent, data-driven framework for modeling housing prices using BNs across three major Spanish cities. By combining structural, spatial, and typological variables from a rich geo-referenced dataset, we constructed interpretable graphical models capable of capturing city-specific pricing logics. Our analysis revealed marked differences between Madrid, Barcelona, and Valencia, not only in the role played by spatial accessibility and housing typology, but also in the structure of dependencies shaping market segmentation. In contrast to black-box methods, our approach supports probabilistic inference, scenario-based reasoning, and sensitivity analysis, offering a layered and explainable understanding of urban housing markets.

Beyond technical accuracy, this work underscores the critical value of interpretability in data-driven housing research. As real estate analytics increasingly inform planning, investment, and policy, the ability to understand and trust model outputs becomes essential. Our study demonstrates that BNs can fill this gap, offering a principled and scalable alternative to opaque models. By integrating insight generation with methodological transparency, this approach moves toward a new standard in urban modeling, one where explanatory clarity supports equitable and accountable housing decisions.

\bibliographystyle{elsarticle-harv} 
 \bibliography{bib}





% \end{thebibliography}

\appendix

\newpage
\section{Appendix: Sensitivity-Colored Bayesian Networks}
\label{appendix:sensitivity_coloring}

\begin{figure}[h]
\centering
\includegraphics[width=0.8\textwidth]{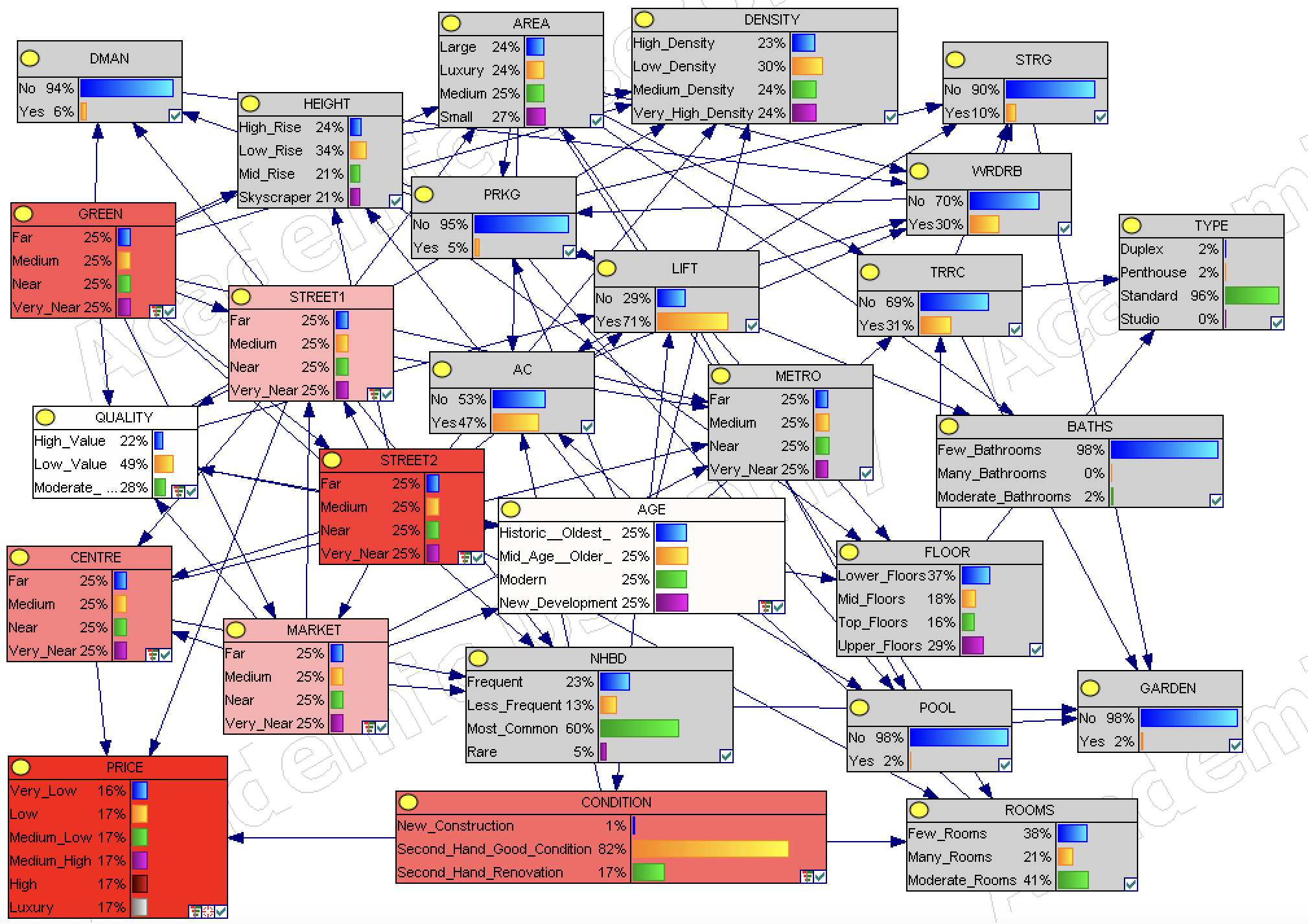}
\caption{Barcelona network with nodes colored by sensitivity to \texttt{PRICE}.}
\label{fig:sens_bcn}
\end{figure}

\begin{figure}
\centering
\includegraphics[width=0.8\textwidth]{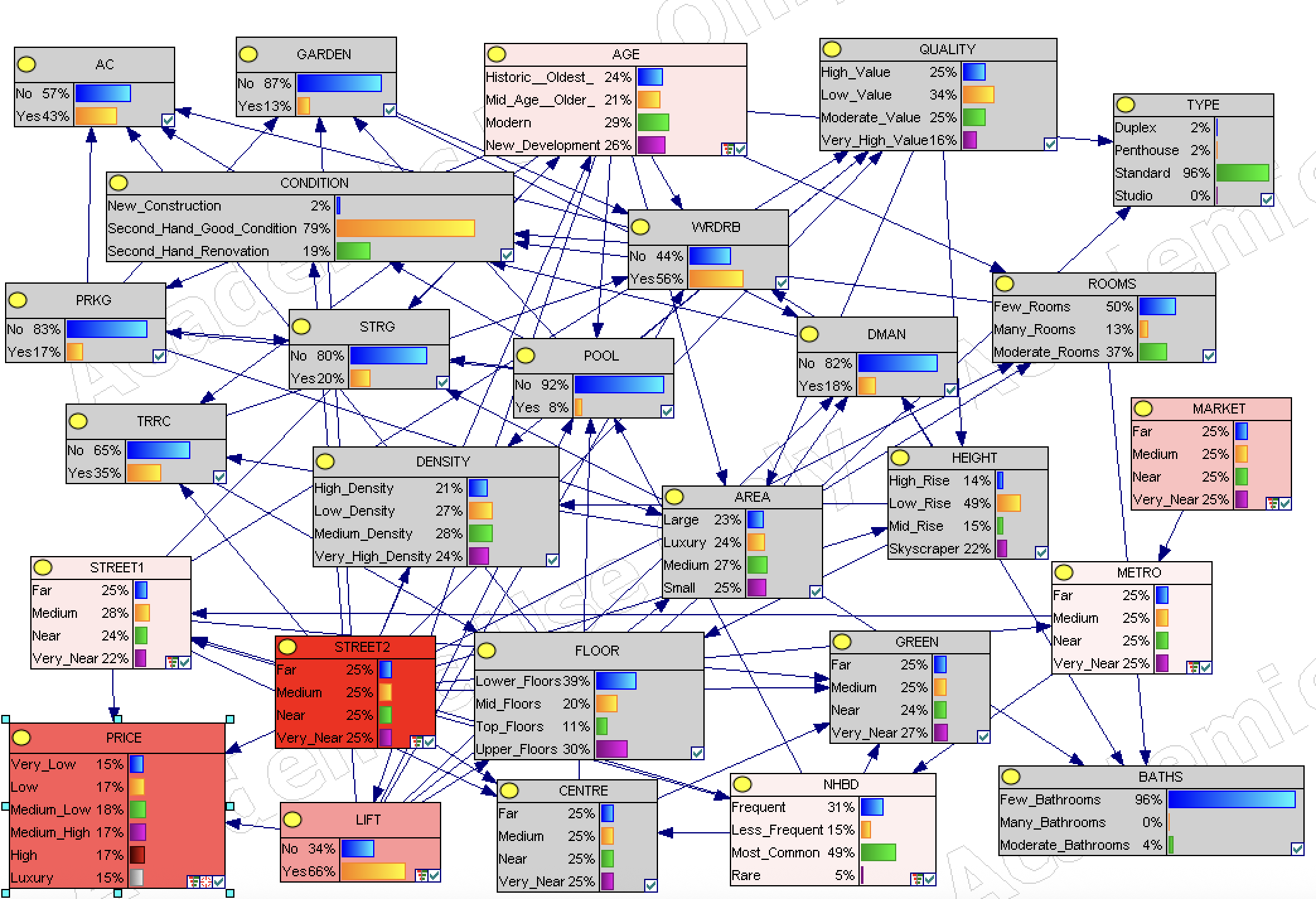}
\caption{Madrid network with nodes colored by sensitivity to \texttt{PRICE}.}
\label{fig:sens_mad}
\end{figure}

\begin{figure}
\centering
\includegraphics[width=0.8\textwidth]{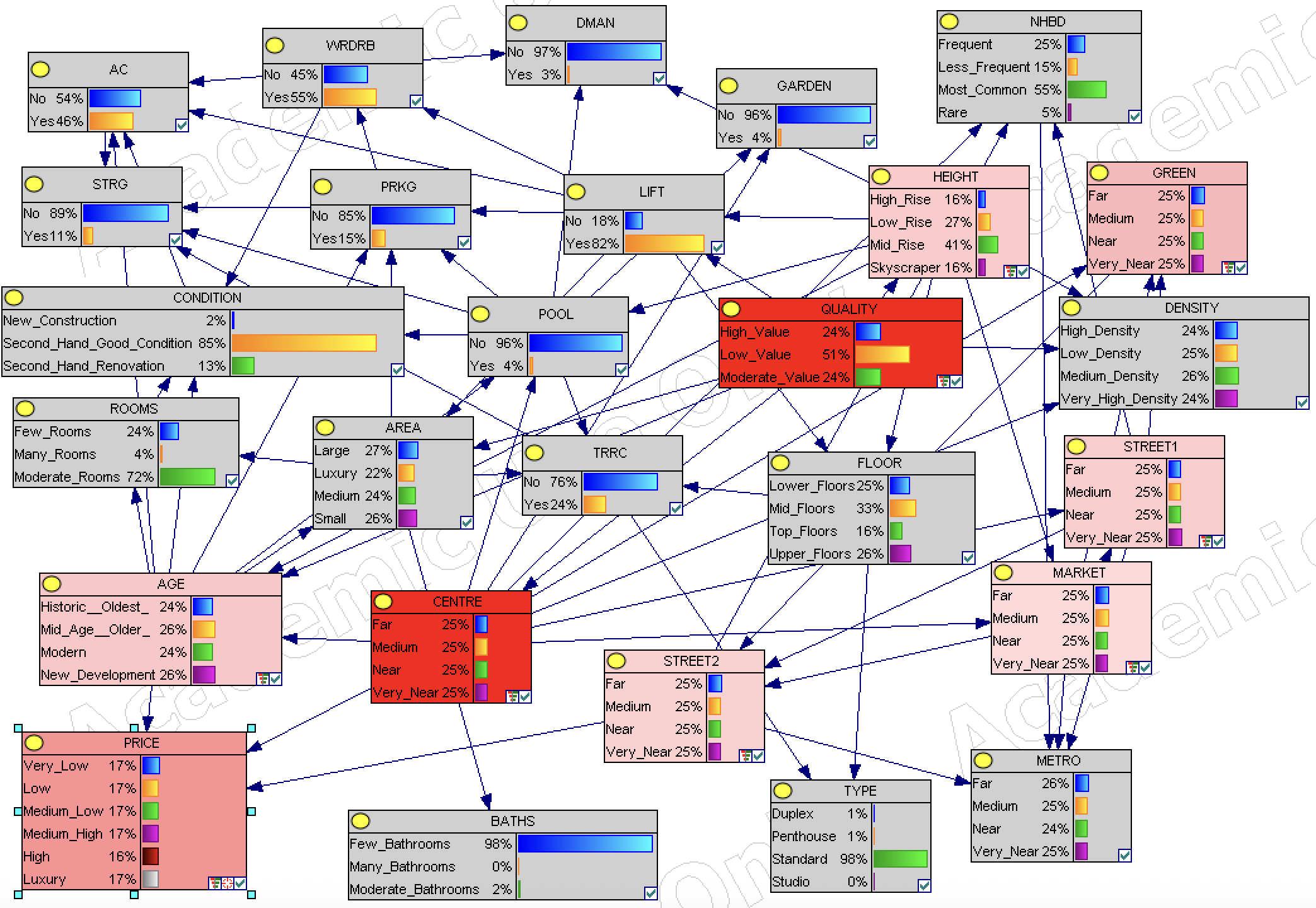}
\caption{Valencia network with nodes colored by sensitivity to \texttt{PRICE}.}
\label{fig:sens_val}
\end{figure}

\end{document}